\pdfoutput=1
\newcommand*{\ATLASLATEXPATH}{latex/}
\documentclass[cernpreprint,texlive=2016,txfonts,UKenglish]{\ATLASLATEXPATH atlasdoc}

\usepackage{\ATLASLATEXPATH atlaspackage}

\usepackage{\ATLASLATEXPATH atlasbiblatex}

\usepackage{\ATLASLATEXPATH atlasphysics}

\graphicspath{{logos/}{figures/}}

\usepackage{vvjj}

\AtlasTitle{Search for diboson resonances with boson-tagged jets in $pp$ collisions at $\sqrt{s}=13$~TeV with the ATLAS detector}

\author{The ATLAS Collaboration}

\AtlasRefCode{EXOT-2016-19}

\PreprintIdNumber{CERN-EP-2017-147}

\AtlasDate{\today}

\AtlasJournal{Phys.\ Lett.\ B.}

\AtlasAbstract{%
Narrow resonances decaying into $WW$, $WZ$ or $ZZ$ boson pairs are searched for in \lumiAll\ fb~$^{-1}$ of proton--proton collision data at a centre-of-mass energy of $\sqrt{s}=13$~TeV recorded with the ATLAS detector at the Large Hadron Collider in 2015 and 2016. The diboson system is reconstructed using pairs of large-radius jets with high transverse momentum and tagged as compatible with the hadronic decay of high-momentum $W$ or $Z$ bosons, using jet mass and substructure properties. The search is sensitive to diboson resonances with masses in the range $1.2$--$5.0$~\TeV{}. No significant excess is observed in any signal region. Exclusion limits are set at the 95\% confidence level on the production cross section times branching ratio to dibosons for a range of theories beyond the Standard Model. Model-dependent lower limits on the mass of new gauge bosons are set, with the highest limit set at 3.5~\TeV{} in the context of mass-degenerate resonances that couple predominantly to bosons.
}

\hypersetup{pdftitle={ATLAS draft},pdfauthor={The ATLAS Collaboration}}

\begin{document}

\maketitle

\clearpage


\clearpage

\section{Introduction}
\label{sec:introduction}

A major goal of the physics programme at the Large Hadron Collider (LHC) is the search 
for new phenomena that may become visible in high-energy proton--proton ($pp$) collisions.
One possible signature of such new phenomena is the production of a heavy resonance with
the subsequent decay into a final state consisting of a pair of vector bosons ($WW$, $WZ$, $ZZ$).
Many models of physics beyond the Standard Model (SM) predict such a signature.
These include extensions to the SM scalar sector as in the 
two-Higgs-doublet model (2HDM)~\cite{Branco:2011iw} that predict new spin-0 resonances, 
composite-Higgs models~\cite{Eichten:2007sx,Sannino:2004qp,Andersen:2011yj} and 
models motivated by Grand Unified Theories~\cite{Pati:1974yy,Georgi:1974sy,Fritzsch:1974nn}
that predict new \wp\ spin-1 resonances,
and warped extra dimensions Randall--Sundrum (RS) models~\cite{Randall:1999ee,ADPS2007,Fitzpatrick}
that predict spin-2 Kaluza--Klein (KK) excitations of the graviton, \GKK.
The heavy vector triplet (HVT)~\cite{Pappadopulo:2014qza,deBlas:2012qp} phenomenological Lagrangian approach 
provides a more model-independent framework for interpretation of spin-1 diboson
resonances.

The search presented here focuses on \TeV{}-scale resonances that decay into
pairs of high-momentum vector bosons which, in turn, decay hadronically.
The decay products of each of those vector bosons are collimated due to the high Lorentz 
boost and are typically contained in a single jet with radius $R = 1.0$. 
While the use of hadronic decays of the vector bosons benefits from the largest
branching ratio (67\% for $W$ and 70\% for $Z$ bosons) amongst the possible final states, 
it suffers from a large background
contamination from the production of multijet events. However, this contamination can
be mitigated with jet substructure techniques that exploit the two-body nature of
\vqq\ decays (with $V = W$ or $Z$).
 
Previous searches for diboson resonances were carried out by the ATLAS and CMS 
collaborations with $pp$ collisions at $\rts = 7$, 8 and 13~\TeV{}.
These include fully leptonic ($\ell\nu \ell\nu$, $\ell\nu \ell\ell$)~\cite{EXOT-2012-11,EXOT-2011-16,EXOT-2013-07,CMS-EXO-12-025}, 
semileptonic ($\nu\nu qq$, $\ell\nu qq$, $\ell\ell qq$)~\cite{EXOT-2016-01,CMS-EXO-13-009,CMS-B2G-16-004}
and fully hadronic ($qqqq$) $VV$~\cite{EXOT-2016-01,CMS-B2G-16-004} final states.
By combining the results of searches in the $\nu\nu qq$, $\ell\nu qq$, $\ell\ell qq$ and
$qqqq$ channels, the ATLAS Collaboration~\cite{EXOT-2016-01}
set a lower bound of 2.60~\TeV{} on the mass of a spin-1 resonance
at the 95\% confidence level, in the context of the HVT model B with $g_V = 3$ 
(described in Section~\ref{sec:models}).
When interpreted in the context of the bulk RS model with a spin-2 KK graviton and
$\kMPl = 1$, this lower mass bound is 1.10~\TeV{}.
The results presented here benefit from an integrated luminosity of \lumiAll~\ifb,
which is an order of magnitude larger than was available for the previous search in the 
fully hadronic final state at $\rts = 13~\TeV$~\cite{EXOT-2016-01}.


\section{Signal models}
\label{sec:models}
The analysis results are interpreted in terms of different models that predict
the production of heavy resonances with either spin 0, spin 1 or spin 2.
In the case of the spin-0 interpretation, a heavy scalar is produced via
gluon--gluon fusion with subsequent decay into a pair of vector bosons.
For this empirical model, the width of the signal in the diboson
mass distribution is assumed to be dominated by the experimental resolution.
The width of a Gaussian distribution characterising the mass resolution
after full event selection 
ranges from approximately 3\% to 2\% as the resonance mass increases from 1.2 to 5.0~\TeV.
The spin-0 model is referred to as the heavy scalar model
in the rest of this Letter.

In the HVT phenomenological Lagrangian model, a new heavy vector triplet (\wp, \zp) is
introduced, with the new gauge bosons degenerate in mass (also denoted by $V'$ in the following). The couplings between those bosons
and SM particles are described in a general manner, thereby allowing a broad class of
models to be encompassed by this approach.
The new triplet field interacts with the Higgs field and thus with the longitudinally polarised
$W$ and $Z$ bosons by virtue of the equivalence theorem~\cite{Cornwall:1974km,Gounaris:1986cr,Chanowitz:1985hj}.
The strength of the coupling to the Higgs field, and thus SM gauge bosons, 
is controlled by the parameter combination $g_V c_H$,
where $c_H$ is a multiplicative constant used to parameterise potential deviations from
the typical strength of triplet interactions to SM vector bosons, taken to be $g_V$. 
Coupling of the triplet field to SM fermions is set by the expression $g^2 c_F/g_V$, where
$g$ is the SM $\mathrm{SU}(2)_\mathrm{L}$ gauge coupling and, like for the coupling to the Higgs field, $c_F$ is 
a multiplicative factor that modifies the typical coupling of the triplet field to fermions.
The HVT model A with $g_V = 1$, 
$c_H \simeq -g^2/g_V^2$ and $c_F \simeq 1$~\cite{Pappadopulo:2014qza} 
is used as a benchmark. In this model, the new triplet field
couples weakly to SM particles and arises from an extension of the SM gauge group.
Branching ratios for $\wp \to WZ$ and $\zp \to WW$ are approximately 2.0\% each.
The intrinsic width $\Gamma$ of the new bosons is approximately 2.5\% of the mass, which results
in observable mass peaks with a width dominated by the experimental resolution.
In this model, the dominant decay modes are into fermion pairs and
searches in the $\ell\ell$ and $\ell\nu$ final states~\cite{EXOT-2016-05,EXOT-2016-06}
provide the best sensitivity.
The calculated production cross section times branching ratio (\sigbr) values for $\wp \to WZ$ with $W$ and $Z$
bosons decaying hadronically are 8.3 and 0.75 fb for $\wp$ masses of 2 and 3 \TeV{}, respectively.
Corresponding values for $\zp \to WW$ are 3.8 and 0.34 fb.

The HVT model B with $g_V = 3$
and $c_H \simeq c_F \simeq 1$~\cite{Pappadopulo:2014qza} 
is used as another benchmark. 
This model describes scenarios in which strong dynamics give rise to the SM Higgs boson
and naturally include a new heavy vector triplet field with electroweak quantum numbers.
The constants $c_H$ and $c_F$ are approximately unity,
and couplings to fermions are suppressed, giving rise to larger branching
ratios ($\sim 50\%$) for either $\wp \to WZ$ or $\zp \to WW$ decays than in model A.
Resonance widths and experimental signatures are similar to those obtained for 
model A and the predicted \sigbr\ values
for $\wp \to WZ$ with hadronic $W$ and $Z$ decays
are 13 and 1.3 fb for $\wp$ masses of 2 and 3 \TeV{}, respectively.
Corresponding values for $\zp \to WW$ are 6.0 and 0.55 fb.

The RS model with one warped extra dimension predicts the existence of spin-2 Kaluza--Klein excitations
of the graviton, with the lowest mode being considered in this search.
While the original RS model~\cite{Randall:1999ee} (often referred to as RS1) 
is constructed with all SM fields confined to a four-dimensional brane (the ``\TeV{} brane''), 
the bulk RS model~\cite{Randall:1999ee,ADPS2007} employed here allows those fields 
to propagate in the extra-dimensional bulk between the \TeV{} brane and the Planck brane.
Although ruled out by precision electroweak and flavour measurements, the RS1 model
is used as a benchmark model to interpret diphoton and dilepton resonance searches 
due to the sizeable \GKK\ couplings to light fermions in that model.
In the bulk RS model, those couplings are suppressed and decays
into final states involving heavy fermions, gauge bosons or Higgs bosons are favoured.
The strength of the coupling depends on \kMPl, where $k$ corresponds to the curvature of
the warped extra dimension, and the effective four-dimensional Planck scale 
$\overline{M}_{\mathrm{Pl}} = 2.4 \times 10^{18}$~\GeV{}.
The cross section and intrinsic width scale as the square of \kMPl.
For the choice $\kMPl = 1$ used in this search, the \sigbr\
values for $\GKK \to WW$ with $W$ decaying hadronically are 0.54 and 0.026 fb for 
$\GKK$ masses of 2 and 3 \TeV{}, respectively.
Corresponding values for $\GKK \to ZZ$ are 0.32 and 0.015 fb. 
In the range of \GKK\ masses considered, the branching ratio to $WW$ ($ZZ$) varies
from 24\% to 20\% (12\% to 10\%) as the mass increases. 
Decays into the $t \bar{t}$ final state dominate with a branching ratio varying
from 54\% to 60\%.
The \GKK\ resonance has a $\Gamma$ value that is approximately 6\% of its mass.


\newcommand{\AtlasCoordFootnote}{%
ATLAS uses a right-handed coordinate system with its origin at the nominal interaction point (IP)
in the centre of the detector and the $z$-axis along the beam pipe.
The $x$-axis points from the IP to the centre of the LHC ring,
and the $y$-axis points upwards.
Cylindrical coordinates $(r,\phi)$ are used in the transverse plane,
$\phi$ being the azimuthal angle around the $z$-axis.
The pseudorapidity is defined in terms of the polar angle $\theta$ as $\eta = -\ln \tan(\theta/2)$.
The rapidity is defined relative to the beam axis as $y = \frac{1}{2} \ln \frac{E+p_z}{E-p_z}$.
Angular distance is measured in units of $\Delta R \equiv \sqrt{(\Delta\eta)^{2} + (\Delta\phi)^{2}}$.}

\section{ATLAS detector}
\label{sec:detector}

The ATLAS experiment~\cite{PERF-2007-01,ATLAS-TDR-19} at the LHC is a multi-purpose particle detector
with a forward-backward symmetric cylindrical geometry and a near $4\pi$ coverage in
solid angle.\footnote{\AtlasCoordFootnote}
It consists of an inner detector for tracking surrounded by a thin superconducting solenoid
providing a \SI{2}{\tesla} axial magnetic field, electromagnetic and hadronic calorimeters, and a muon spectrometer.
The inner detector covers the pseudorapidity range $|\eta| < 2.5$.
It consists of silicon pixel, silicon microstrip, and transition radiation tracking detectors.
A new innermost pixel layer~\cite{ATLAS-TDR-19} inserted at a radius of 3.3~cm has been used since 2015.
Lead/liquid-argon (LAr) sampling calorimeters provide electromagnetic (EM) energy measurements
with high granularity.
A hadronic (steel/scintillator-tile) calorimeter covers the central pseudorapidity range ($|\eta| < 1.7$).
The end-cap and forward regions are instrumented with LAr calorimeters
for both the EM and hadronic energy measurements up to $|\eta| = 4.9$.
The muon spectrometer surrounds the calorimeters and features
three large air-core toroidal superconducting magnet systems with eight coils each.
The field integral of the toroids ranges between \num{2.0} and \SI{6.0}~Tm
across most of the detector. The muon spectrometer includes a system of precision tracking 
chambers and fast detectors for triggering.
A two-level trigger system~\cite{Aaboud:2016leb} is used to select events.
The first-level trigger is implemented in hardware and uses a subset of the detector information
to reduce the accepted rate to at most \SI{100}{\kilo\hertz}.
This is followed by a software-based trigger level that reduces the accepted event rate 
to \SI{1}{\kilo\hertz} on average.


\section{Data and simulation}
\subsection{Data}
\label{sec:data}

The data for this analysis were collected during the LHC
$pp$ collision running at $\sqrt{s} = 13$~\TeV{} in 2015 and 2016.
Events must pass a trigger-level requirement of having at least one large-radius jet with 
transverse energy $\ET > 360$~\GeV{} in 2015 and $\ET > 420$~\GeV{} in 2016, where the
jet is reconstructed using the \antikt\ algorithm~\cite{antikt} 
with a radius parameter of 1.0.
Those thresholds correspond to the 
lowest-\ET, unprescaled large-radius jet triggers for each of the two data-taking periods.
After requiring that the data were collected during stable beam conditions and the
detector components relevant to this analysis were functional,
the integrated luminosity of the sample amounts to \lumiFifteen ~\ifb\ and \lumiSixteen ~\ifb\ of
$pp$ collisions in 2015 and 2016, respectively.


\subsection{Simulation}
\label{sec:simulation}

The search presented here uses simulated Monte Carlo (MC) event samples
to optimise the selection criteria,
to estimate the acceptance for different signal processes, and 
to validate the experimental procedure described below.
However, it does not rely on MC event samples 
to estimate the background contribution from SM processes.

Signal events for the heavy scalar model~\cite{Barger:2007im} 
were produced at next-to-leading-order via the gluon--gluon fusion mechanism 
with \POWHEGBOX v1~\cite{Alioli:2010xd,Alioli:2008tz} 
using the CT10 parton distribution function (PDF) set~\cite{ct10}.
Events were interfaced with \PYTHIAV{v8.186}~\cite{pythia8} for parton showering 
and hadronisation using the CTEQ6L1 PDF set~\cite{Pumplin:2002vw} and 
the AZNLO set of tuned parameters (later referred to as tune)~\cite{STDM-2012-23}. 
The width of the heavy scalar is negligible compared to the experimental resolution.

In the case of the HVT and RS models, events were produced at leading order (LO) with the
\MGMCatNLO v2.2.2 event generator~\cite{madgraph} using the NNPDF23LO PDF set~\cite{Ball:2012cx}.
To study the sensitivity of the spin-2 resonance search to production from quark--antiquark or gluon--gluon initial states as well as to different vector-boson
polarisation states, events were generated with JHUGen v5.6.3~\cite{Gao:2010qx}
and the NNPDF23LO PDF set.
For these signal models, the event generator was interfaced with 
\PYTHIAV{v8.186} for parton showering and hadronisation with
the A14 tune~\cite{ATL-PHYS-PUB-2014-021}.
The \GKK\ samples are normalised according to calculations from Ref.~\cite{Oliveira:2014kla}.
In all signal samples, the $W$ and $Z$ bosons are longitudinally polarised.

Multijet background events were generated with \PYTHIAV{v8.186} with the NNPDF23LO PDF set
and the A14 tune.
Samples of $W$+jets and $Z$+jets events were generated with 
\HERWIGpp v2.7.1~\cite{Bahr:2008pv} using the CTEQ6L1 PDF set
and the UEEE5 tune~\cite{Gieseke:2012ft}.

For all MC samples, 
charm-hadron and bottom-hadron decays were handled by {\textsc{EvtGen}} v1.2.0~\cite{Lange:2001uf}.
Minimum-bias events generated using \PYTHIA8 were added to the hard-scatter interaction in such a way as to reproduce the effects of additional $pp$ interactions in
each bunch crossing during data collection (pile-up).
An average of 23 pile-up interactions are observed in the data in addition to the hard-scatter interaction.
The detector response was simulated with \GEANT~4~\cite{geant, SOFT-2010-01}
and the events were processed with the same reconstruction software as for the data.


\section{Event reconstruction and selection}
\subsection{Reconstruction}
\label{sec:reconstruction}

The selection of events relies on the identification and reconstruction
of electrons, muons, jets, and missing transverse momentum. 
Although the analysis primarily relies on jets, other
particle candidates are needed to reject events that are included in 
complementary searches for diboson resonances.

The trajectories of charged particles are reconstructed using measurements in the
inner detector.
Of the multiple $pp$ collision vertices reconstructed from the available
tracks in a given event, a primary vertex is selected as the one with the
largest $\sum \pt^2$, where the sum is over all tracks with 
transverse momentum $\pt > 0.4$~\GeV{} 
that are associated with the vertex. 
Tracks that are consistent with the primary vertex may be identified 
as electron or muon candidates.
Electron identification is based on matching tracks to energy clusters
in the electromagnetic calorimeter and relying on the longitudinal and
transverse shapes of the electromagnetic shower. 
Electron candidates are required to satisfy the ``medium'' identification criterion~\cite{ATLAS-CONF-2016-024} and to pass the ``loose'' track-based
isolation~\cite{ATLAS-CONF-2016-024}.
Muon identification relies on matching tracks in the inner detector to muon spectrometer tracks or track segments. 
Muon candidates must also satisfy the ``medium''
selection criterion~\cite{PERF-2015-10} and the ``loose'' track 
isolation~\cite{PERF-2015-10}.

Large-radius jets (hereafter denoted large-$R$ jets) are reconstructed from 
locally calibrated clusters of energy deposits
in calorimeter cells~\cite{PERF-2011-03} with the \antikt\ clustering 
algorithm using a radius parameter $R = 1.0$.
Jets are trimmed~\cite{Krohn2010} to minimise the impact of pile-up
by reclustering the constituents of each jet with the \kt\ algorithm~\cite{Ellis:1993tq}
into smaller $R = 0.2$ subjets and removing those subjets with 
$\pt^{\text{subjet}} / \,\pt^{\text{jet}} < 0.05$, where $\pt^{\text{subjet}}$
and $\pt^{\text{jet}}$ are the transverse momenta of the subjet and original
jet, respectively.
The clustering and trimming algorithms use the FastJet package~\cite{Cacciari:2011ma}.
Calibration of the trimmed jet \pt\ and mass is described in Ref.~\cite{ATL-PHYS-PUB-2015-033}.

The large-$R$ jet mass is computed using measurements from the calorimeter and
tracking systems~\cite{ATLAS-CONF-2016-035} according to

\begin{equation}
\label{eqn:jetmass}
\nonumber
  \mj = w^{\text{cal}} m^{\text{cal}} + w^{\text{trk}}\frac{\pt}{\pt^{\text{trk}}} m^{\text{trk}} ,
\end{equation}

where 
$\pt^{\text{trk}}$ is the transverse momentum of the jet evaluated
using only charged-particle tracks associated with the jet,
$m^{\text{cal}}$ and $m^{\text{trk}}$ are the masses computed using calorimeter and
tracker measurements, and
$w^{\text{cal}}$ and $w^{\text{trk}}$ are weights inversely proportional to the
square of the resolution of each of the corresponding mass terms.
Ghost association~\cite{Cacciari:2007fd} is performed to associate tracks 
to the jets before the trimming procedure is applied.
In this method, tracks are added with an infinitesimally small
momentum as additional constituents in the jet reconstruction.
Tracks associated with the jets are required to have $\pt > 0.4$~\GeV{} and
satisfy a number of quality criteria based on the number of measurements in the silicon pixel and microstrip detectors;
tracks must also be consistent with originating from the primary vertex~\cite{ATLAS-CONF-2016-035}.
Including information from the tracking system provides improved mass resolution, especially at high jet \pt, due to the relatively coarse angular resolution of the calorimeter.

The magnitude of the event's missing transverse momentum (\met)
is computed from the vectorial sum of calibrated electrons, muons,
and jets in the event~\cite{PERF-2014-04}. 
For this computation and the rejection of non-collision
background discussed below, jets are reconstructed from
topological clusters 
using the \antikt\ algorithm with a radius parameter $R = 0.4$ and are required to
satisfy $\pt > 20$ \GeV{} and $|\eta| < 4.9$. Calibration of those jets
is described in Ref.~\cite{PERF-2016-04}.
The \met\ value is corrected using tracks associated with the primary
vertex but not associated with electrons, muons or jets.


\subsection{Selection}
\label{sec:selection}

Events used in complementary searches for diboson resonances in different
final states are removed, in anticipation of a future combination. 
Accordingly, events are rejected if they contain any electron or muon
with $\pt > 25$ \GeV{} and $|\eta| < 2.5$.
Furthermore, events with $\met > 250$ \GeV{} are rejected.

Events with jets that are likely to be due to
non-collision sources, including calorimeter noise, beam halo and
cosmic rays, are removed~\cite{ATLAS-CONF-2015-029}. 
Events are required to contain at least two large-$R$ jets
with $|\eta| < 2.0$ (to guarantee a good overlap with the tracking acceptance)
and mass $\mj > 50$~\GeV{}.
The leading (highest \pt) large-$R$ jet must have $\pt > 450$~\GeV{} and
the subleading (second highest \pt) large-$R$ jet must have $\pt > 200$~\GeV{}.
The invariant mass of the dijet system formed by these two jets must
be $\mjj > 1.1$~\TeV{} to avoid inefficiencies due to the minimum jet-\pt requirements
and to guarantee that the trigger requirement is fully efficient.
Only jets in this system are considered in the rest of this Letter.
Events passing the above requirements are said to pass the event ``preselection''.

Further kinematic requirements are imposed to suppress background from multijet production.
The rapidity separation between the leading and subleading jets (identified with subscripts
1 and 2 in the following) must be sufficiently 
small,  ${|\Delta y| = |y_1 - y_2| < 1.2}$, which is particularly aimed at suppressing
$t$-channel dijet production. The $\pt$ asymmetry between the two jets 
$A = (\ptone - \pttwo)\,/\,(\ptone + \pttwo)$ must be smaller than 0.15
to remove events where one jet is poorly reconstructed.

Jets must be consistent with originating from hadronic decays
of $W$ or $Z$ bosons. Discrimination against background jets
inside a mass window including the $W/Z$ mass is based on the 
variable \dtwo, which is defined as a ratio of two-point to three-point energy correlation 
functions 
that are based on the energies of and pairwise angular distances between the jet's constituents~\cite{Larkoski:2014gra,Larkoski:2015kga}. 
This variable is optimised with parameter $\beta = 1$ to distinguish between 
jets originating from a single parton and those coming from the two-body decay of a heavy 
particle. A detailed description of the optimisation can be found in 
Refs.~\cite{PERF-2015-03,ATL-PHYS-PUB-2015-033}.
The boson-tagging criteria---the jet-mass window size and maximum \dtwo\ value---are 
simultaneously optimised to  
achieve the maximal background-jet rejection for a fixed $W$ or $Z$
signal-jet efficiency of 50\%. The optimisation uses signal jets from
simulated $W' \to WZ \to qqqq$ events and background jets from simulated multijet events,
and depends on the jet \pt\ to account for varying resolution as a function of jet \pt.
The size of the $W$ ($Z$) mass window varies from 22 (28)~\GeV{} 
near $\pt = 600$~\GeV{} to 40 (40)~\GeV{} at $\pt \ge 2500$~\GeV{}
and the maximum \dtwo\ value varies from 1.0 to 2.0 as the jet \pt\ increases.
An event is tagged as a candidate $WW$ ($ZZ$) event if both jets are within
the $W$ ($Z$) mass window. It can also be tagged as a candidate $WZ$ event
if the lower- and higher-mass jets are within the $W$ and $Z$ mass windows, respectively.
Because the mass windows are relatively wide and overlap, jets may pass both $W$- and $Z$-tagging requirements.

To specifically suppress gluon-initiated jets, the number of tracks associated
with each jet must satisfy $\ntrk < 30$.
The tracks used must have $\pt > 0.5$ \GeV{} and $|\eta| < 2.5$, as well as
originate from the primary vertex.

The above set of selection criteria constitutes the signal region (SR) definition.
Figure~\ref{fig:Signal-efficiency} illustrates the kinematic acceptance times 
selection efficiency (\acceff) at different selection stages 
for simulated heavy scalar resonances, heavy gauge bosons and KK gravitons 
decaying to the $WW$ final state. 
Similar \acceff\ values are obtained in the $WZ$ final state for the
HVT model and in the $ZZ$ final state for the heavy scalar and bulk RS models.
Multijet background events are suppressed with a rejection factor of approximately
$2 \times 10^5$, as determined from simulation.
The figure shows that, among the different selection criteria described above,
the boson tagging reduces the signal \acceff\ the most. However, this particular
selection stage provides the most significant suppression of the dominant
multijet background.

\begin{figure}[hpt!]
\begin{center}
  \subfloat[Heavy scalar model]{
  \includegraphics[width=0.50\textwidth]{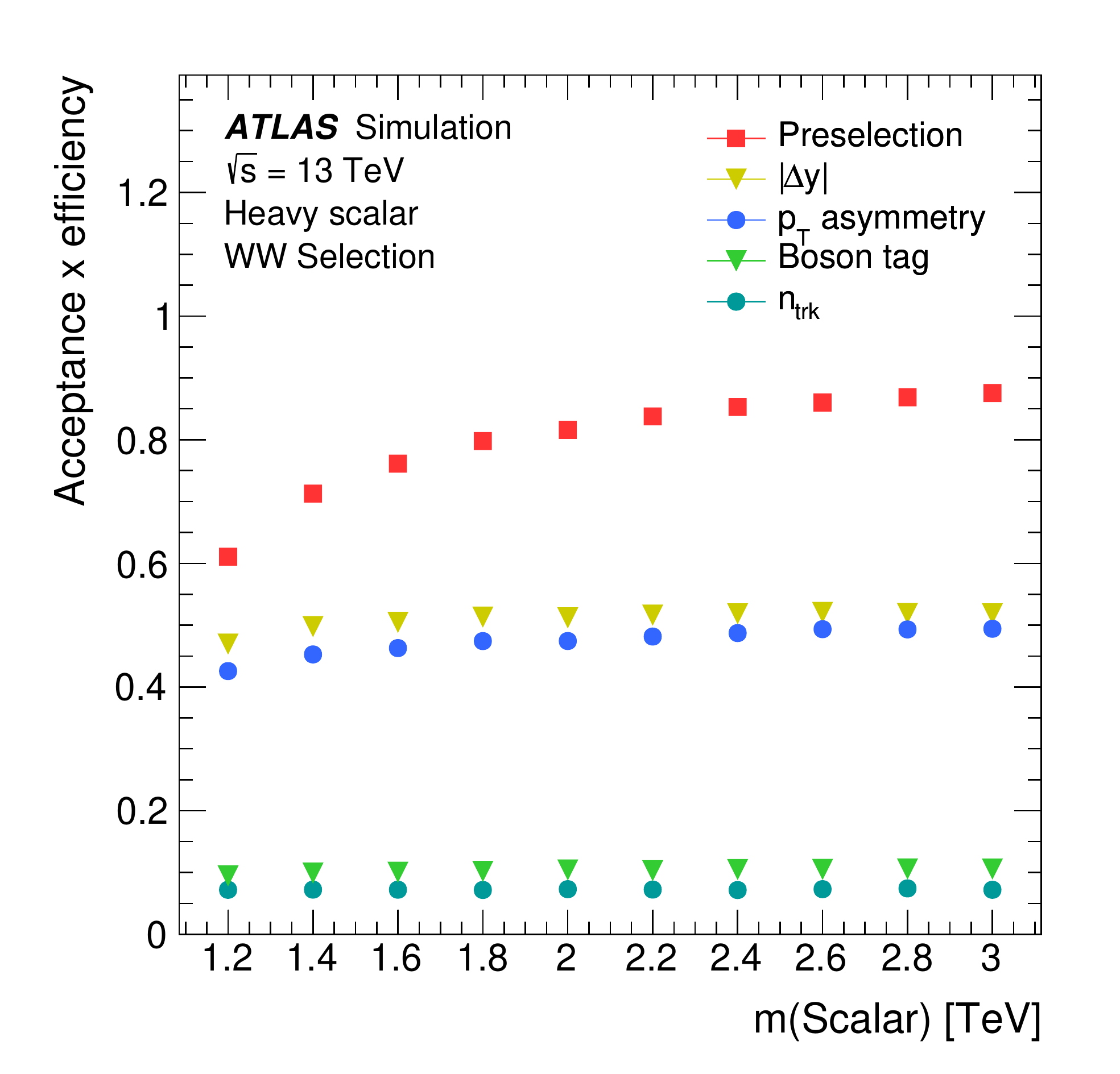}
  \label{fig:WW-NWH}
  } 
  \subfloat[HVT model A, $g_V = 1$]{
  \includegraphics[width=0.50\textwidth]{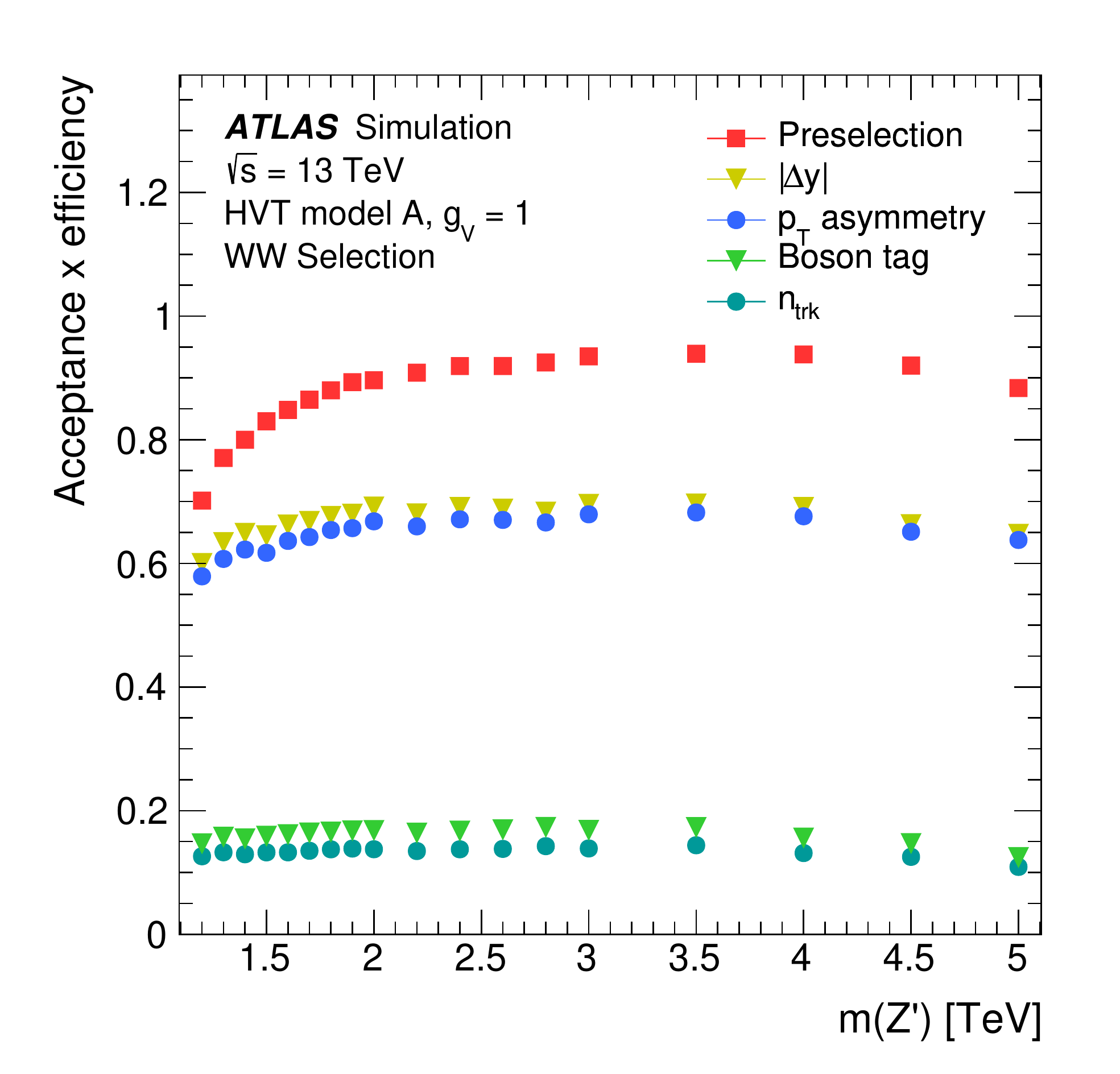}
  \label{fig:WW-HVT}
  } \\
  \subfloat[Bulk RS model, \kMPl = 1]{
  \includegraphics[width=0.50\textwidth]{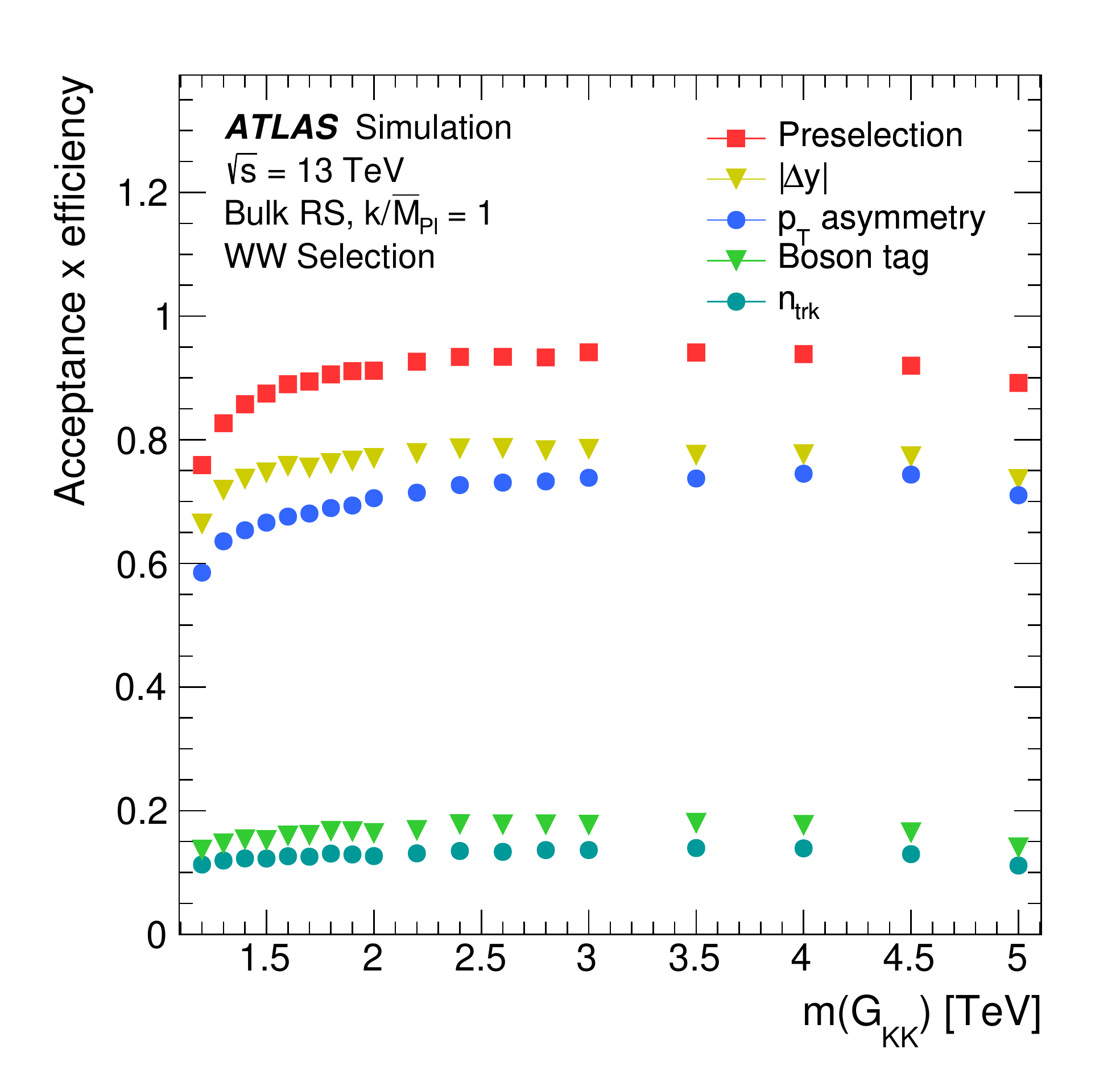}
  \label{fig:WW-RS}
  }
\end{center}
\vspace{-15 pt}
\caption{Signal acceptance times efficiency as a function of resonance mass for
(a) Scalar $\to WW$ in the heavy scalar model, (b) $\zp \to WW$
in the HVT model, and (c) $\GKK \to WW$ in the bulk RS model.
The acceptance times efficiency is shown at successive stages of selection with the final stage ($\ntrk$) 
corresponding to the signal region.}
\label{fig:Signal-efficiency}
\end{figure} 

Table~\ref{tab:efficiency} summarises the \acceff\ values for a number
of models at resonance mass values of 2 and 3~\TeV{} for the $WW$ final state;
similar results are obtained for the other diboson final states. In the case of the
bulk RS model, the KK gravitons are mostly produced via gluon-induced processes
and decay into longitudinally polarised $W$ bosons. 
The polarisation affects the angular separation and momentum sharing between the
decay products in the $W \to qq$ decay and thus affects the boson-tagging efficiency.
To test the impact of the polarisation, the \acceff\ values are evaluated with 
dedicated signal MC samples initiated by only gluons or quarks, and with $W$ bosons
either fully longitudinally polarised or transversely polarised.
Significant differences in the signal \acceff\ are observed, as can be seen
in Table~\ref{tab:efficiency}, and these may need to be taken into account in
reinterpretations of the results presented in this Letter.
Little dependence is observed on the resonance mass.
Differences in \acceff\ for gluon- and quark-initiated production arise primarily 
from differences in the acceptance for selection on the jet $|\eta|$ of the two
leading jets and their rapidity separation.
The boson-tagging efficiency for transversely polarised $W$ bosons is approximately half 
that for longitudinally polarised $W$ bosons and does not depend appreciably on 
the heavy-resonance production mechanism.
In the case of quark-initiated production, \acceff\ is similar for longitudinally and transversely polarised $W$ bosons, as the reduction in kinematic acceptance is approximately compensated by an increase in boson-tagging efficiency.
In the case of gluon-initiated production, both kinematic acceptance and
boson-tagging efficiency favour longitudinally polarised $W$ bosons.

\begin{table}[tb]
\caption{Signal acceptance times efficiency for resonances with masses of
2 and 3~\TeV{} decaying into the $WW$ final state in different models.
The first three rows correspond to values obtained with the nominal signal MC samples
described in Section~\ref{sec:simulation} and the values in the last four rows are 
obtained with the alternate signal MC samples generated with JHUGen.
\label{tab:efficiency}
}
\begin{center}
\begin{tabular}{lcc}
\hline\hline
Model / process & \multicolumn{2}{c}{Acceptance $\times$ efficiency} \\
                & $m = 2$ \TeV  & $m = 3$ \TeV \\
\hline
Heavy scalar            & ~~7.3\%     & ~~7.2\%   \\
HVT model A, $g_V = 1$  &  13.8\%     &  13.9\%  \\
Bulk RS, $\kMPl = 1$    &  12.7\%     &  13.6\%  \\
\hline
$gg \to \GKK \to WW$ (longitudinally polarised $W$)       &  12.3\%  &  13.4\% \\
$gg \to \GKK \to WW$ (transversally polarised $W$)        & ~~1.8\%  & ~~1.9\% \\
$q\bar{q} \to \GKK \to WW$ (longitudinally polarised $W$) & ~~5.4\%  & ~~5.4\% \\
$q\bar{q} \to \GKK \to WW$ (transversally polarised $W$)  & ~~5.2\%  & ~~5.8\% \\
\hline\hline
\end{tabular}
\end{center}
\end{table}

\subsection{Validation}
\label{sec:validation}

In addition to the nominal SR, several validation regions (VRs) are defined
to check the analysis procedure and estimate some of the
sources of systematic uncertainty.

The definitions of the signal and validation regions are summarised in Table~\ref{tab:selection}.
A check of the statistical approach described in Section~\ref{sec:background} is performed
in the three different sideband validation regions. These correspond to the same selection 
as for the signal region except for requiring the jet mass to be in one of two sidebands. 
Both jet masses must
be below the $W$ boson mass with $50 < \mj < 60-72$ \GeV{} (low--low sideband),
or above the $Z$ boson mass with $106-110 < \mj < 140$ \GeV{} (high--high sideband),
or with one jet mass belonging to the low-mass range and the other to the high-mass range
(low--high sideband).
These mass ranges are chosen to have no overlap with the \pt-dependent $W$
and $Z$ mass windows applied to define the signal regions. 
The \pt-dependent mass windows imply a range of $60-72$ \GeV{} for the upper edge
of the lower sideband and $106-110$ \GeV{} for the lower edge of the higher sideband.

\begin{table}[t]
\caption{Event selection requirements and definition of the different regions used in the 
 analysis. Different requirements are indicated for the highest-\pt (leading) jet with
 index 1 and the second highest-\pt (subleading) jet with index 2. 
 The jet mass boundaries applied in the definition of the sideband validation regions
 depend on the jet \pt.
}
\begin{center}
\begin{tabular}{l  l}
\hline\hline
 Signal region      & Veto non-$qqqq$ channels: \\
                    & ~~~~~~ No $e$ or $\mu$ with $\pt > 25$ \GeV{} and $|\eta| < 2.5$ \\
                    & ~~~~~~ $\met < 250$ \GeV{} \\
                    & Event preselection: \\
                    & ~~~~~~ $\geq 2$ large-$R$ jets with $|\eta| < 2.0$ and $\mj > 50$ \GeV{} \\
                    & ~~~~~~ $\ptone > 450$ \GeV{} and $\pttwo > 200$ \GeV{} \\
                    & ~~~~~~ $\mjj > 1.1$ \TeV{} \\
                    & Topology and boson tag: \\
                    & ~~~~~~ $|\Delta y| = |y_1 - y_2| < 1.2$ \\
                    & ~~~~~~ $A = (\ptone - \pttwo) \, / \, (\ptone + \pttwo) < 0.15$ \\
                    & ~~~~~~ Boson tag with \dtwo\ variable and $W$ or $Z$ mass window \\
                    & ~~~~~~ $\ntrk < 30$ \\
\midrule\midrule
Low--low sideband       & Same selection as for signal region, except: \\
~~~~~ validation region & ~~~~~ $50 < m_1 < 60-72$ \GeV{} and $50 < m_2 < 60-72$ \GeV{} \\
\midrule
High--high sideband     & Same selection as for signal region, except: \\
~~~~~ validation region & ~~~~~ $106-110 < m_1 < 140$ \GeV{} and $106-110 < m_2 < 140$ \GeV{} \\
\midrule
Low--high sideband      & Same selection as for signal region, except: \\
~~~~~ validation region & ~~~~~ $50 < m_1 < 60-72$ \GeV{} and $106-110 < m_2 < 140$ \GeV{}, or \\
                        & ~~~~~ $106-110 < m_1 < 140$ \GeV{} and $50 < m_2 < 60-72$ \GeV{} \\
\midrule
$V$+jets validation region 
                    & Veto non-$qqqq$ channels (see above) \\
                    & $V$+jets selection: \\
                    & ~~~~~~ $\geq 2$ large-$R$ jets with $|\eta| < 2.0$\\
                    & ~~~~~~ $\ptone > 600$ \GeV{} and $\pttwo > 200$ \GeV{} \\
                    & ~~~~~~ Boson tag with \dtwo\ variable only applied to leading jet \\
\hline\hline
\end{tabular}
\label{tab:selection} 
\end{center}
\end{table}

A $V$+jets validation region is defined primarily to compare the observed
and simulated $V$+jets event yields as a function of the number of tracks 
associated with the large-$R$ jets and thereby derive an uncertainty in
the efficiency for the \ntrk\ requirement.
There is no attempt at using this validation region to constrain the $V$+jets
contribution to the signal regions as the total background there is estimated from
an empirical fit to the dijet mass distribution.
The $V$+jets validation region requires the presence of at least two
large-$R$ jets with $|\eta| < 2.0$. 
The leading jet must satisfy $\pt > 600$ \GeV{} and the subleading jet $\pt > 200$ \GeV{}.
A higher minimum \pt\ requirement is imposed on the leading jet than in the nominal event
selection to obtain a sample with higher average leading jet \pt\ that better corresponds
to the jet \pt\ values probed in the search.
Finally, the leading jet must pass the boson-tagging requirements based on the
\dtwo\ variable only (i.e. the jet mass is not included in the tagging); 
no boson tagging is applied to the subleading jet.
The resulting event sample in this validation region is approximately an order of magnitude larger than
the samples selected in the different signal regions.
Figure~\ref{fig:Vjets-mass} shows the leading jet mass distribution in
the range  $50 <\mj <150$~\GeV{} for events
in this $V$+jets validation region for $\ntrk < 30$ and $\ntrk \ge 30$.
A clear contribution of $W/Z$ events is visible for $\ntrk < 30$ but it is much less
apparent for $\ntrk \ge 30$, supporting the use of an upper limit on the number of tracks
in the signal region.

To establish the efficiency in data of the $\ntrk < 30$ selection, the leading-jet mass distribution is analysed in eight multiplicity subsamples, covering $0 \leq n_\mathrm{trk} \leq 39$ in groups of five tracks each. 
Events originating from $W$+jets and $Z$+jets processes are modelled using
a double-Gaussian distribution with the shape parameters determined from simulation, 
while background events not originating from
$V$+jets processes are fit to data independently in each subsample using a fourth-order
polynomial (denoted ``Fit bkd.'' in Figure~\ref{fig:Vjets-mass}). 
The relative normalisation in each $\ntrk$ bin is
controlled by a function which has a scaling parameter, allowing a
variation in the track efficiency. 
The relative $W$ and $Z$ boson event contributions are fixed to the prediction
from the simulation but the total $W + Z$ event normalisation is determined in the fit.
A small upward shift in the $W/Z$ boson peak position is observed as $\ntrk$ increases, which is
well modelled by the simulation.
An overall data-to-simulation scale factor of $1.03 \pm 0.05$ is extracted for the $\ntrk$ requirement 
per $V$ jet. As this factor is consistent with unity, no correction is applied.

\begin{figure}[t]
\begin{center}
  \subfloat[Leading-jet mass for $\ntrk < 30$]{
  \includegraphics[width=0.50\textwidth]{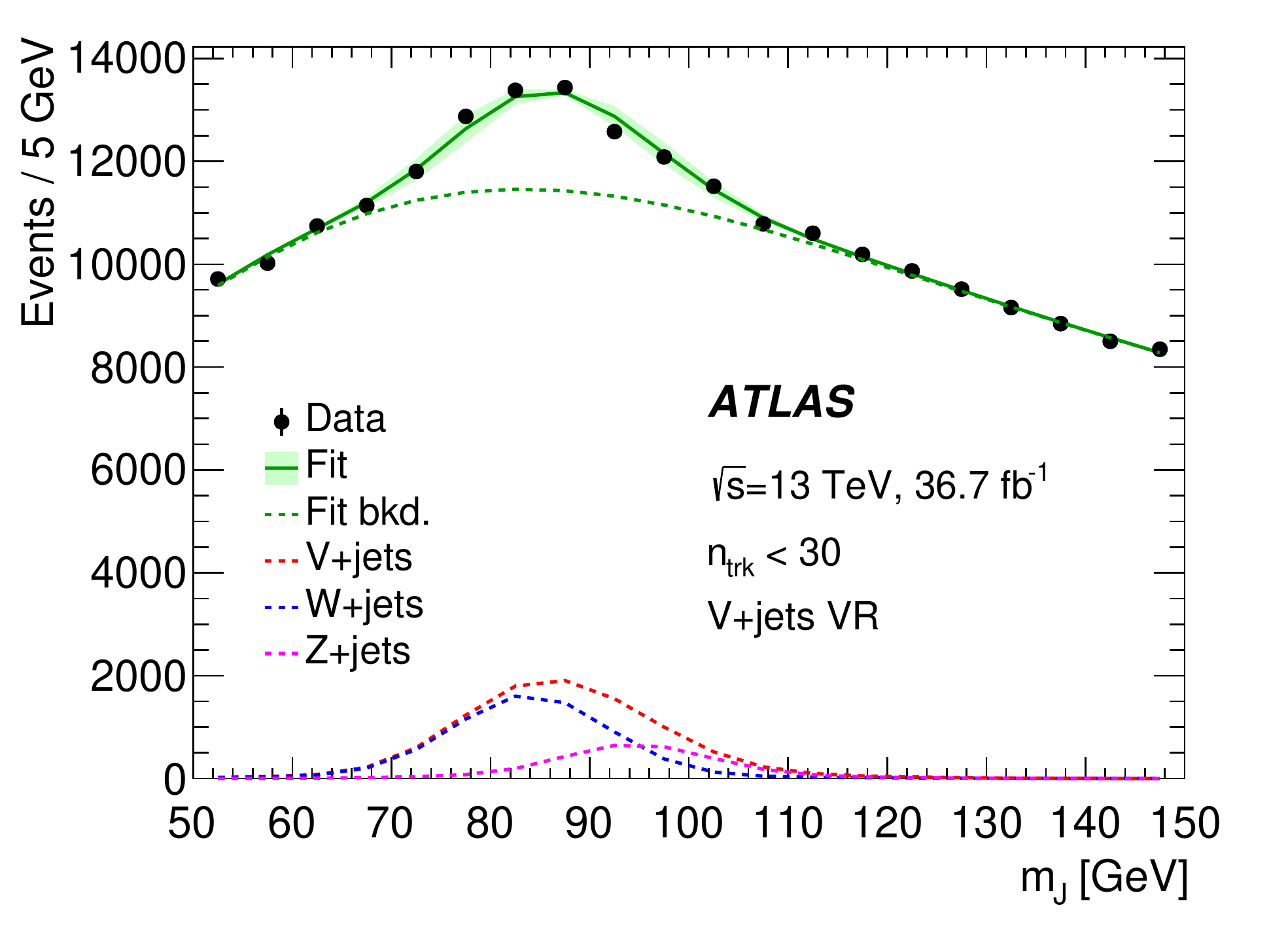}
  \label{fig:jetmass-ntrklt30}
  } 
  \subfloat[Leading-jet mass for $\ntrk \geq 30$]{
  \includegraphics[width=0.50\textwidth]{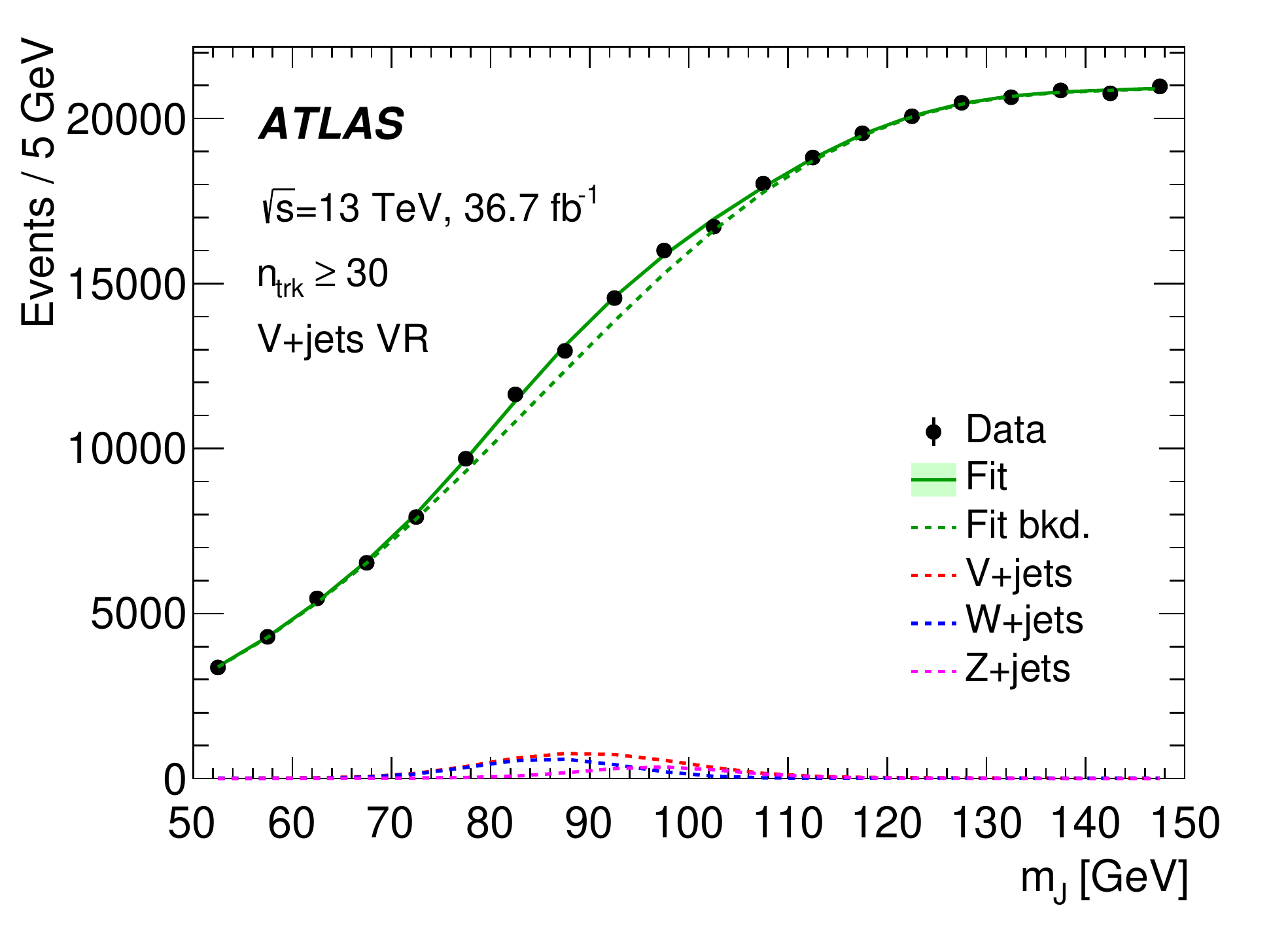}
  \label{fig:jetmass-ntrkge30}
  }
\end{center}
\vspace{-15 pt}
\caption{Leading-jet mass distribution for data in the $V$+jets validation region
 for two different ranges of track multiplicity after boson tagging based only on
 the \dtwo\ variable.
 The result of fitting to the sum of functions for the $V$+jets and background events
 is also shown, and described in the text. The error band around the fit result
 corresponds to the uncertainty in the jet mass scale.}
\label{fig:Vjets-mass}
\end{figure} 

\section{Background parameterisation}
\label{sec:background}

The search for diboson resonances is performed by looking for narrow peaks above the smoothly falling $\mjj$ distribution expected in the SM.  
This smoothly falling background mostly consists of SM multijet events. Other SM processes, including diboson, $W/Z$+jets and $t\bar{t}$ production, amount to about $15\%$ of the total background. They are also expected to have smoothly falling invariant mass distributions, although not necessarily with the same slope.
The background in this search is estimated empirically from a binned maximum-likelihood
fit to the observed $\mjj$ spectrum in the signal region. 
The following parametric form is used:  

\begin{equation}
\label{eqn:dijet12}
 \frac{\mathrm{d}n}{\mathrm{d}x} = p_1 (1-x)^{p_2-\xi p_3} x^{-p_3} ,
\end{equation}

where $n$ is the number of events, $x = \mjj/\sqrt{s}$, $p_1$ is a normalisation factor,
$p_2$ and $p_3$ are dimensionless shape parameters, and
$\xi$ is a constant chosen to 
remove the correlation between $p_2$ and $p_3$ in the fit.
The latter is determined by repeating the fit with different $\xi$ values.
The observed $\mjj$ distribution in data is histogrammed 
with a constant bin size of 100~\gev{} and the parametric form above is fit
in the range ${1.1 < \mjj < 6.0}$~\TeV{}. 
Only $p_2$ and $p_3$ are allowed to vary in the fit since $p_1$
is fixed by the requirement that the integral of ${\mathrm{d}n}/{\mathrm{d}x}$ equals
the number of events in the distribution. 
This function has been successfully used in previous iterations of this analysis~\cite{EXOT-2016-01}.
Other functional forms were tested and no significant improvement in the fit quality was observed.

The ability of the parametric shape in Eq.~(\ref{eqn:dijet12}) to model the expected 
background distribution is tested in the three 
background-enriched sideband validation regions defined in Table~\ref{tab:selection}. 
The results of the fits to data are shown in Figure~\ref{bkg:fig:VR} 
along with the $\chi^2$ per degree of freedom (DOF).
Bins with fewer than five events are grouped with bins that contain 
at least five events to compute the number of degrees of freedom.
The fit model is found to provide a good description of the data in all of the VRs.

\begin{figure}[pt!]
\begin{center}
\subfloat[Low--low validation region]{
\includegraphics[scale=\scfigtwob]{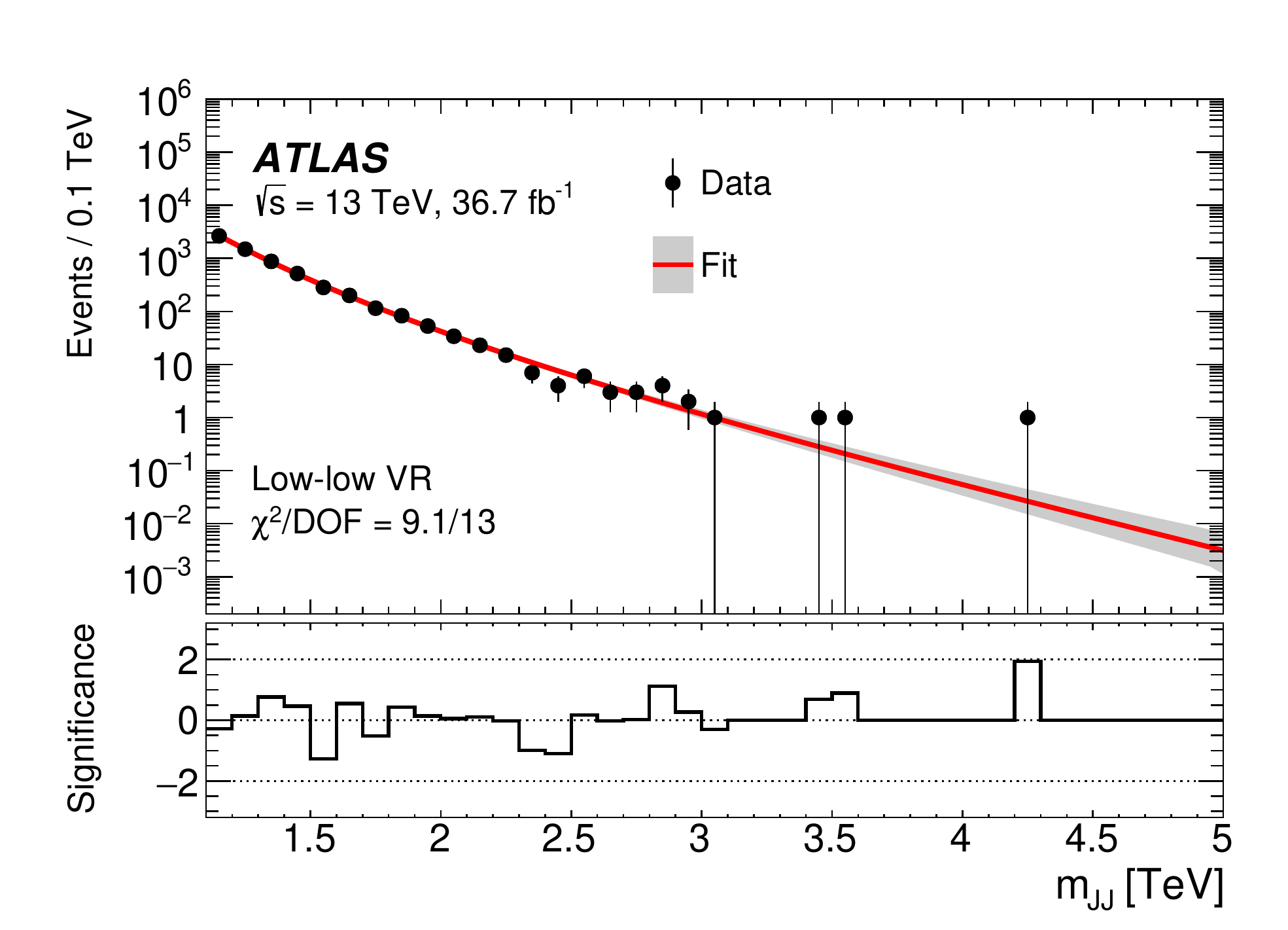}\label{bkg:fig:LL_data}
}
  \subfloat[Low--high validation region]{
\includegraphics[scale=\scfigtwob]{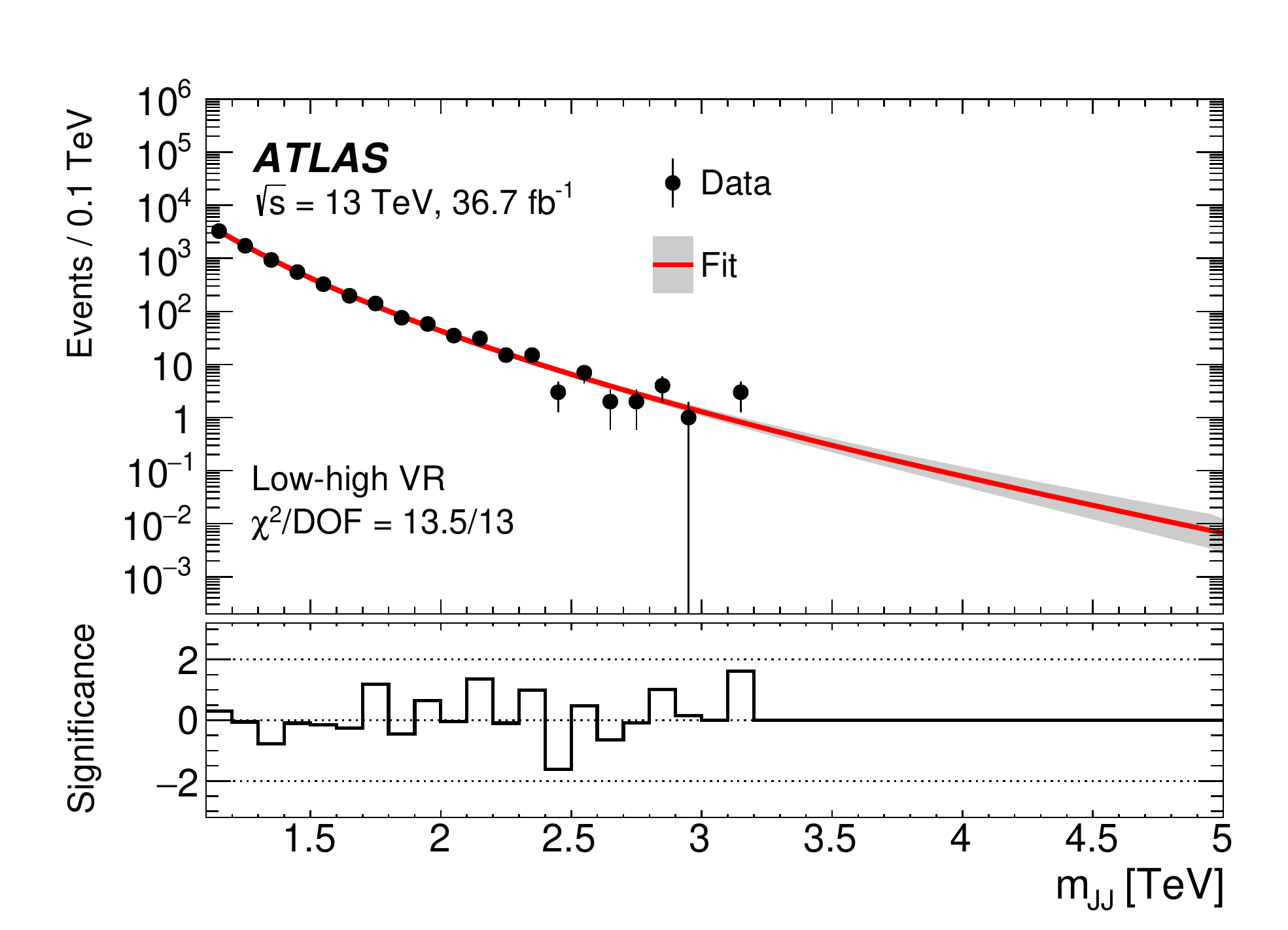}\label{bkg:fig:LH_data}
} \\
 \subfloat[High--high validation region]{
\includegraphics[scale=\scfigtwob]{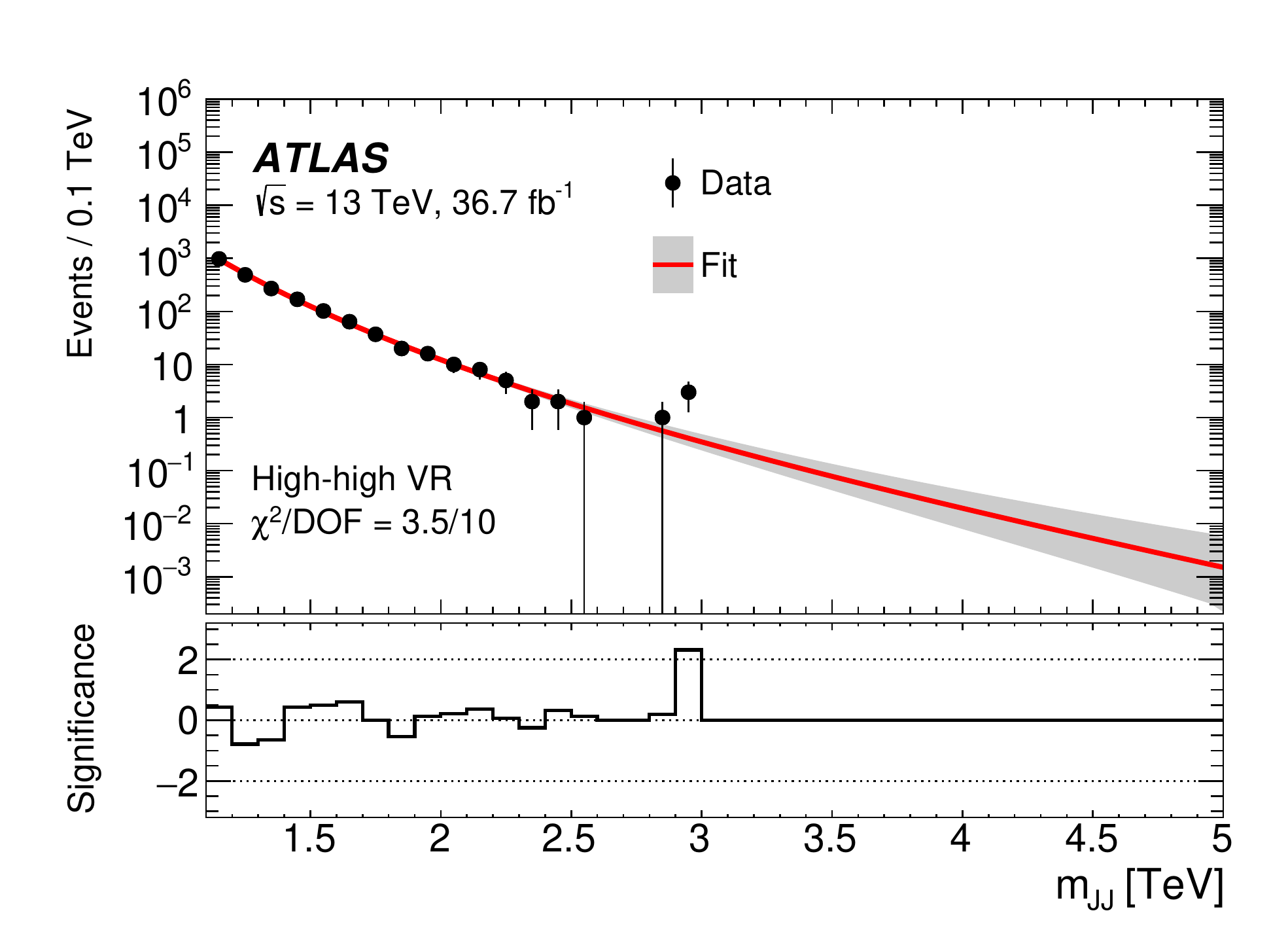}\label{bkg:fig:HH_data}
}
\end{center}
\vspace{-20 pt}
\caption{Dijet mass distributions for data in the sideband validation regions. The solid lines correspond to the result of the fit and the shaded bands represent the uncertainty in the background expectation. 
The lower panels show the significance of the observed event yield relative to the background fits taking their uncertainties into account as described in Ref.~\cite{Choudalakis:2012}.
}
\label{bkg:fig:VR}
\end{figure} 


A profile likelihood test following Wilks' theorem~\cite{Wilks:1938dza} is used 
to determine if including an additional parameter in the background model is necessary. 
Using the simulated multijet background with the sample size expected for the 2015+2016 dataset,
as well as large sets of pseudo-experiments, Eq.~(\ref{eqn:dijet12}) is found to be 
sufficient to describe the data.
Possible additional uncertainties due to the choice of background model are assessed 
by performing signal-plus-background fits (also called spurious-signal tests) 
to the data in the sideband validation regions, where a signal contribution is expected 
to be negligible. The background is modelled with Eq.~(\ref{eqn:dijet12}) and 
the signal is modelled using resonance mass distributions from simulation. 
The signal magnitude obtained in these background-dominated regions 
is less than 25\% of its statistical uncertainty at any of the resonance masses
considered in this search. Therefore, no additional uncertainty is assigned.


\section{Systematic uncertainties}
\label{sec:systematics}

Systematic uncertainties in the signal yield and $\mjj$ distribution are assessed, and expressed as additional nuisance parameters in the statistical analysis, as described in Section \ref{sec:statistics}. The dominant sources of uncertainty in the signal modelling arise from uncertainties in the large-$R$ jet energy and mass calibrations, affecting the jet $p_{\mathrm{T}}$, mass and \dtwo\ values. The correlations between the uncertainties in these jet variables are investigated by calculating the resulting uncertainties in the yield at a variety of signal mass points for three different configurations: ``strong'', with all three variables fully correlated; ``medium'', with $p_{\mathrm{T}}$ and $\mj$ correlated, whilst the \dtwo\ is uncorrelated; and ``weak'', with all three variables fully uncorrelated. The ``medium'' configuration is chosen as it results in the most conservative (largest) uncertainty in the yield. 

Uncertainties in the modelling of the jet energy scale (JES), jet mass scale (JMS) and \dtwo\ scale are evaluated using track-to-calorimeter double ratios between data and MC simulation~\cite{PERF-2012-02}. This method introduces additional uncertainties from tracking. Uncertainties associated with track reconstruction efficiency, impact parameter resolution, tracking in dense environments, rate for fake tracks and sagitta biases are included. The size of the total correlated JES (JMS) uncertainty varies with jet $p_{\mathrm{T}}$ and is approximately $3\%$ ($5\%$) per jet for the full signal mass range. The uncorrelated scale uncertainty in \dtwo\ also varies with jet $p_{\mathrm{T}}$ and is approximately $3\%$ per jet for the full signal mass range.

Uncertainties in the modelling of jet energy resolution (JER), jet mass resolution (JMR) and \dtwo\ resolution are assessed by applying additional smearing of the jet observables according to the uncertainty in their resolution measurements~\cite{PERF-2012-02,ATL-PHYS-PUB-2015-033}. For the JER a $2\%$ absolute uncertainty is applied per jet, and to mass and \dtwo\ relative uncertainties of $20\%$ and $15\%$ are applied per jet, respectively. The response of the \dtwo\ requirement is not strictly Gaussian and therefore the RMS of the observed distribution is taken as an approximation of the nominal width.
There are sufficient dijet data to derive jet-related uncertainties up to jet \pt\ values of 3~\TeV{}~\cite{ATLAS-CONF-2017-063}.

The efficiency of the $\ntrk < 30$ requirement in data and MC simulation is evaluated in the $V$+jets VR defined in Section~\ref{sec:validation}. 
The $\ntrk$ efficiency scale factor is predominantly extracted using jets with $\pt\approx650$~GeV, whereas signal jets in the analysis extend to $\pt\ge1$~\TeV{}. Examining the distribution of the number of tracks associated with jets as a function of jet $\pt$ reveals similar increasing trends in data and MC simulation. However, the average track multiplicity in the simulation is $3\%$ larger at high \pt. Combining the 5\% track multiplicity scale uncertainty with the $\ntrk$ modelling uncertainty leads to a total 6\% uncertainty per tagged jet in the efficiency of the $\ntrk$ requirement.
The uncertainty from the trigger selection is found to be negligible, as the minimum requirement on the dijet invariant mass of 1.1~\TeV{} guarantees that the trigger is fully efficient. 

Uncertainties affecting the signal prediction are as follows.
The uncertainty in the combined 2015+2016 integrated luminosity is 3.2\%. It is derived, following a methodology similar to that detailed in Ref.~\cite{Aaboud:2016hhf}, from a calibration of the luminosity scale using $x$--$y$ beam-separation scans performed in August 2015 and May 2016. 
Theoretical uncertainties in the signal prediction are accounted for via their impact
on the signal acceptance. The uncertainty associated with PDFs at high $Q^2$ values is modelled by taking the envelope formed by the largest deviations produced by the error sets of three PDF sets, as set out by the PDF4LHC group~\cite{Butterworth:2015oua}. For the HVT model, the uncertainty ranges from 0.5\% to 6\% depending on the mass being tested, while a constant 0.5\% uncertainty is determined in the case of the heavy scalar and bulk RS models. Uncertainties arising from the choice of A14 tuning parameters are covered by producing samples with variations of the tuning parameters describing initial-state radiation, final-state radiation, and multi-parton interactions. The uncertainty in the signal acceptance is then evaluated at MC generator level, before boson tagging or $n_{\mathrm{trk}}$ cuts, resulting in a constant uncertainty of 3\% for the HVT model and 5\% for the heavy scalar and bulk RS models.


\section{Results}
\label{sec:results}
\subsection{Background fit}
\label{sec:Fit}
The fitting procedure described in Section~\ref{sec:background} is applied to the data passing the $WW$, $WZ$ and $ZZ$ selections described in Section~\ref{sec:selection}, and resulting dijet mass distributions are shown in Figure~\ref{fig:Data-results}. 
The mass spectra obtained in combined $WW+WZ$ and $WW+ZZ$ SRs are also shown. 
A total of 497, 904, 618, 980, and 904 events are found in the $WW$, $WZ$, $ZZ$, $WW+WZ$, and $WW+ZZ$ SRs. 
Approximately 20\% of events are included in all three regions: $WW$, $WZ$ and $ZZ$. The requirements of the $WW$ ($ZZ$) SR are satisfied by 47\% (57\%) of the events in the $WZ$ SR.
The fitted background functions shown, labelled ``Fit'', are evaluated in bins between 1.1 and 6.0~\TeV. No events are observed beyond $3.1$~\TeV{}. The dijet mass distributions in all signal regions are described well by the background model over the whole range explored.

\begin{figure}[hpt!]
\begin{center}
  \subfloat[$WW$ selection]{
\includegraphics[scale=\scfigtwob]{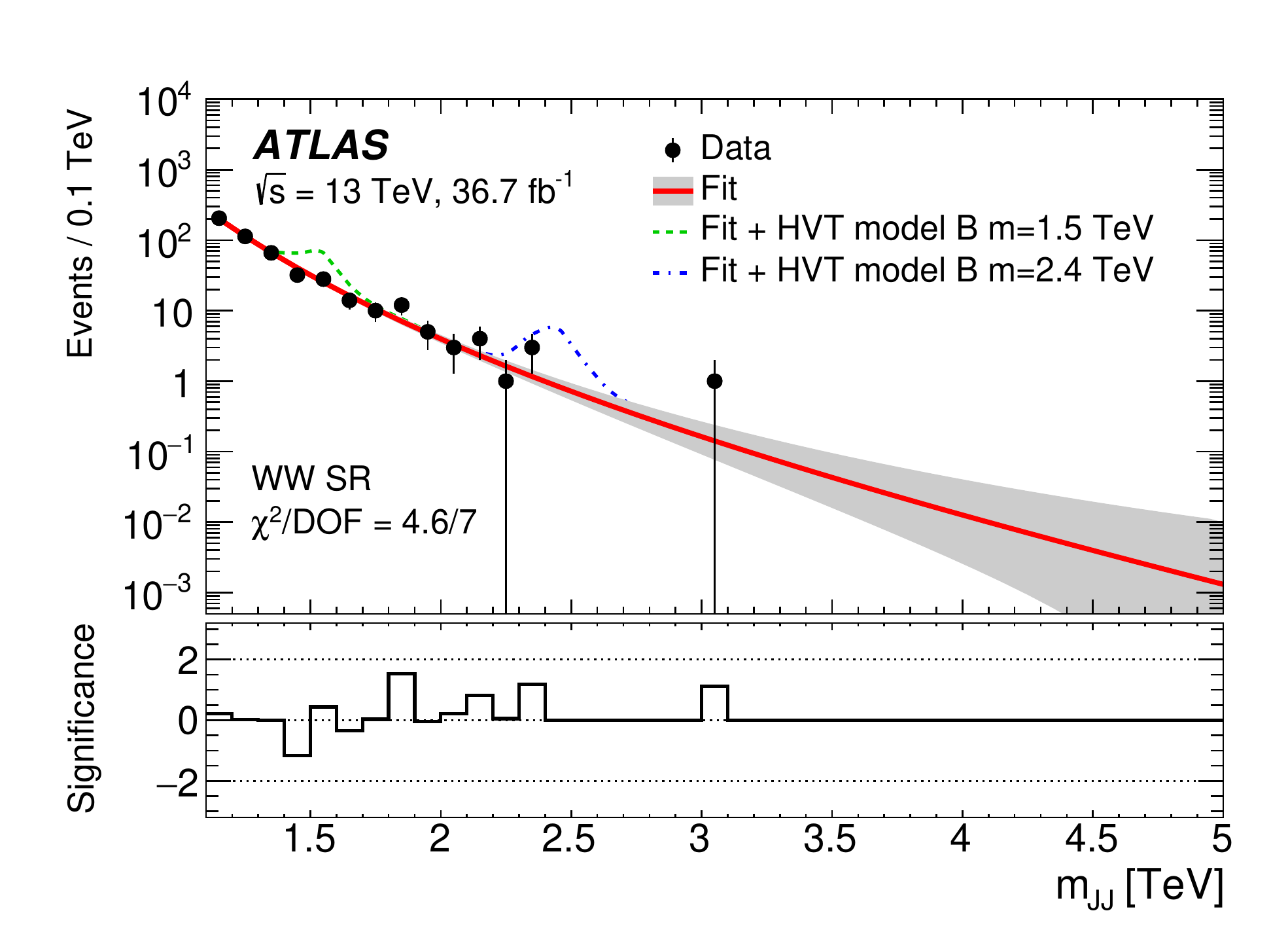}
\label{fig:WW-data}
} 
  \subfloat[$WZ$ selection]{
\includegraphics[scale=\scfigtwob]{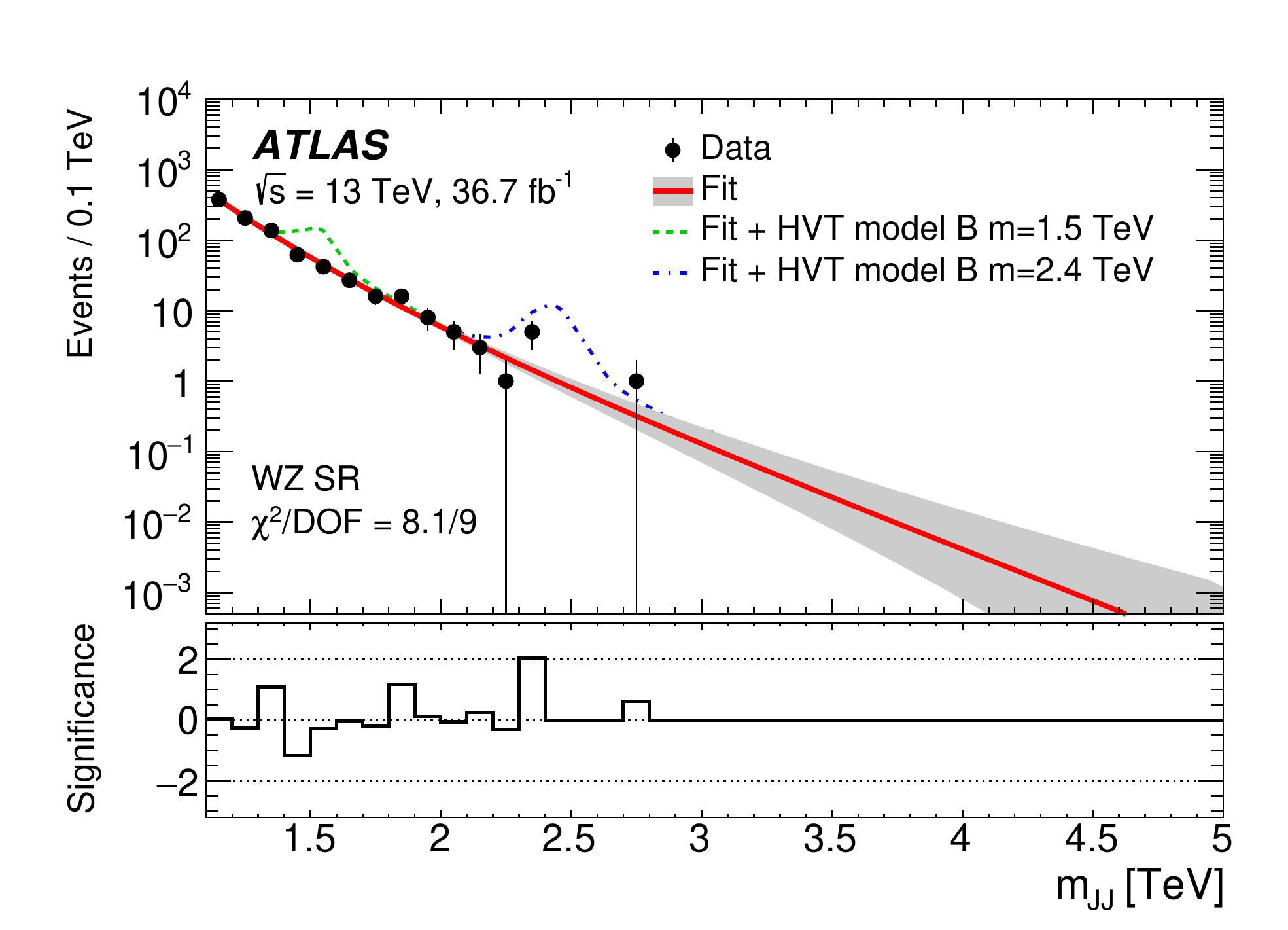}
\label{fig:WZ-data}
} \\
  \subfloat[$ZZ$ selection]{
\includegraphics[scale=\scfigtwob]{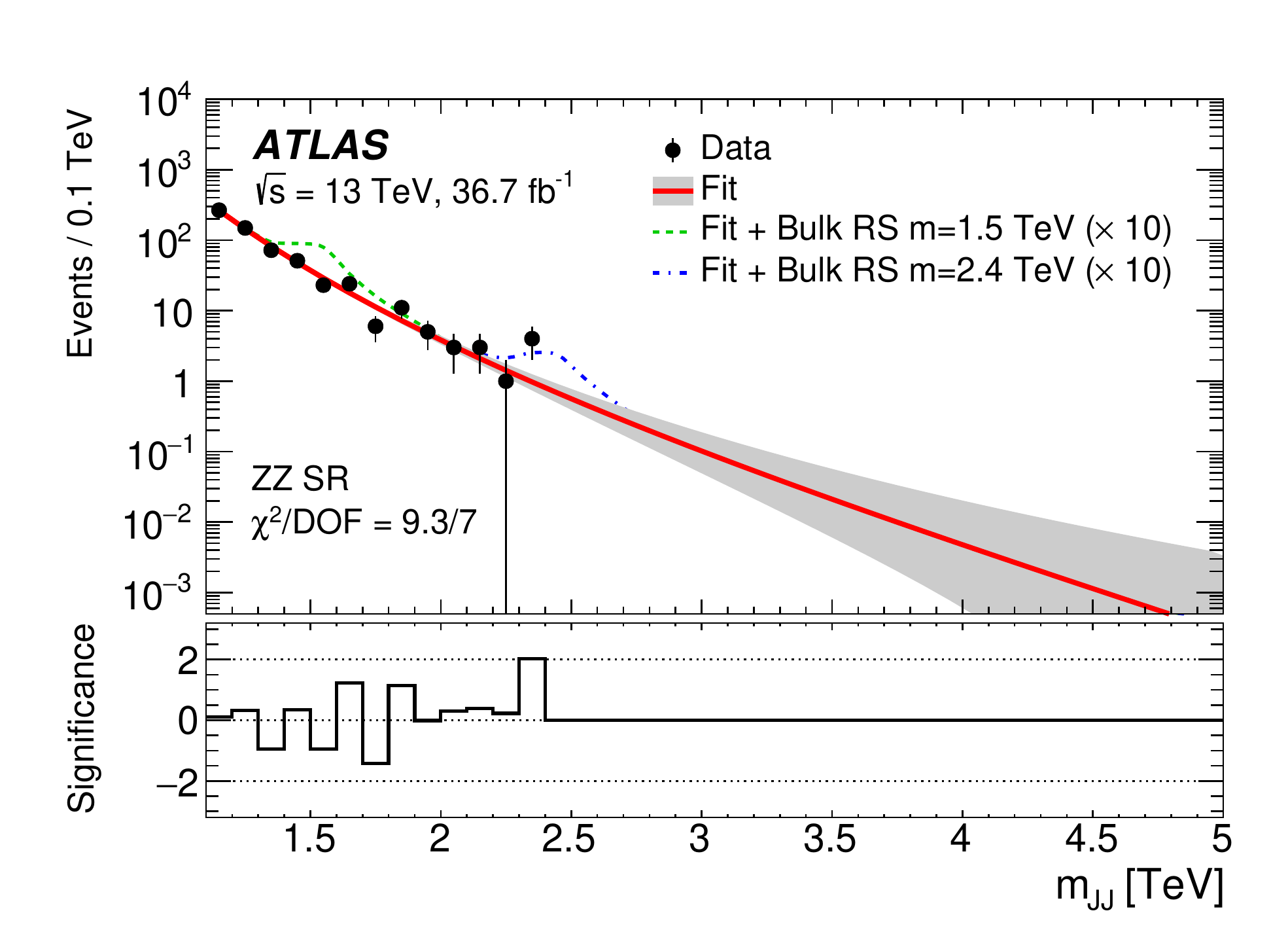}
\label{fig:ZZ-data}
} 
  \subfloat[$WW+WZ$ selection]{
\includegraphics[scale=\scfigtwob]{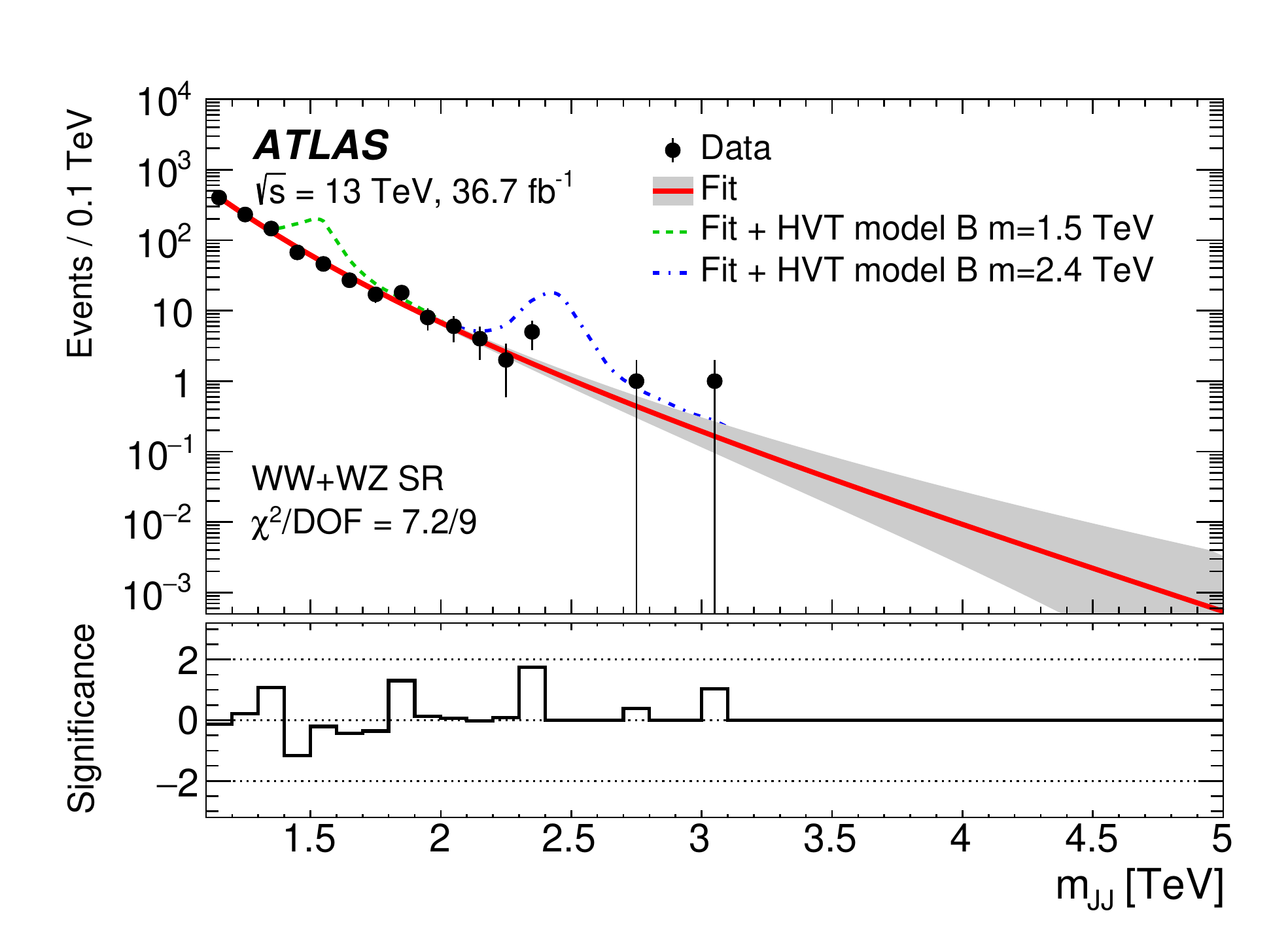}
\label{fig:WZorWW-data}
} \\
  \subfloat[$WW+ZZ$ selection]{
\includegraphics[scale=\scfigtwob]{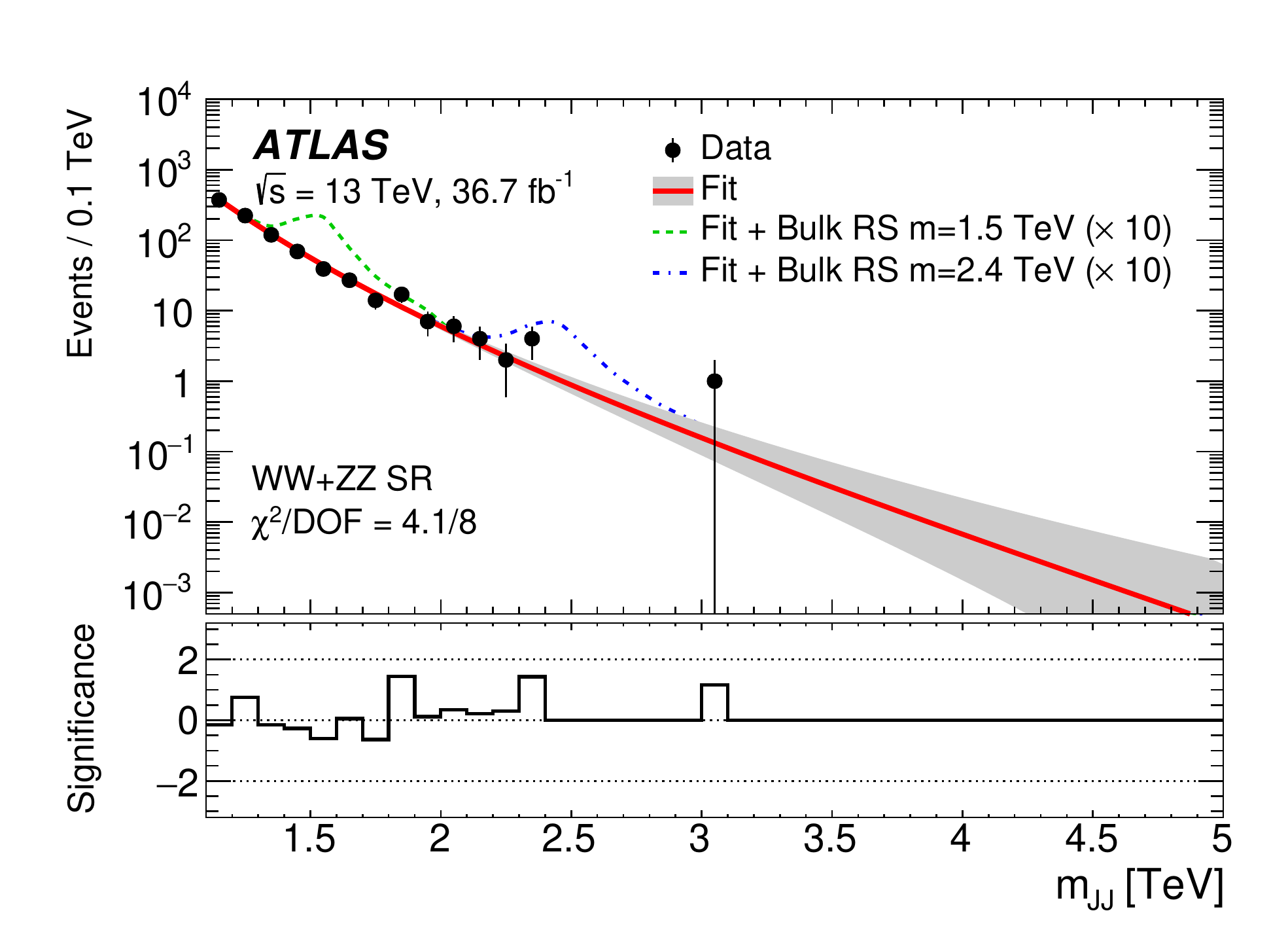}
\label{fig:WWorZZ-data}
}
\end{center}
\vspace{-20 pt}
\caption{Dijet mass distributions for data in the (a) $WW$, (b) $WZ$, and (c) $ZZ$ signal regions, as well as in the combined (d) $WW+WZ$ and (e) $WW+ZZ$ signal regions. The solid lines correspond to the result of the fit and the shaded bands represent the uncertainty in the background expectation. The lower panels show the significance of the observed event yield relative to the background fits.
 Expected signals are shown for the HVT model B with $g_V = 3$ and the bulk RS model
 with $\kMPl = 1$. The predictions for {\GKK} production are multiplied by a factor of 10.
 The lower panels show the significance of the observed event yield relative to the background fits taking their uncertainties into account as described in Ref.~\cite{Choudalakis:2012}.
 \label{fig:Data-results}
}
\end{figure} 

As a test of the background model, the fit is also performed on dijet mass distributions obtained 
with no boson tagging applied but with weights corresponding to the probability for each jet to 
satisfy the boson tagging requirements. This probability is derived from the data as a function 
of the jet \pt\ and the resulting fits are consistent with the nominal background fits within 
uncertainties. The use of untagged data allows to validate the model with a sufficiently large 
number of data events up to dijet masses of 6~\TeV.

\subsection{Statistical analysis}
\label{sec:statistics}

The final results are interpreted using a frequentist statistical analysis. The parameter of interest is taken to be the signal strength, $\mu$, defined as a scale factor to the number of signal events predicted by the new-physics model being tested. A test statistic $\lambda(\mu)$, based on a profile likelihood ratio \cite{Cowan:2010js} is used to extract information about $\mu$ from a maximum-likelihood fit of the signal-plus-background model to the data. The likelihood model is defined as
\begin{equation}
\label{eqn:lhood}
\nonumber
  {\cal L} = \prod_i P_{\mathrm{pois}}( \nobsi | \nexpi ) \times G(\alpha) \times {\cal N}(\theta) ,
\end{equation}
where $P_{\mathrm{pois}}( \nobsi | \nexpi )$ is the Poisson probability to observe $\nobsi$ events
in dijet mass bin $i$
if $\nexpi$ events are expected, $G(\alpha)$ are a series of Gaussian probability density functions modelling the systematic uncertainties, $\alpha$, related to the shape of the signal, and ${\cal N}(\mathbf{\theta})$ is a log-normal distribution for the nuisance parameters, $\theta$, which model the systematic uncertainty in the signal normalisation. 
The expected number of events is the bin-wise sum of those expected for the signal and background:
  $n_{\mathrm{exp}} = n_{\mathrm{sig}} + n_{\mathrm{bg}}$. The expected number of background events in bin $i$, $n_{\mathrm{bg}}^i$,
is obtained by integrating ${\mathrm d}n/{\mathrm d}x$ obtained from Eq.~(\ref{eqn:dijet12}) over that bin.
Thus, $n_{\mathrm{bg}}$ is a function of the background parameters $p_1$, $p_2$, and $p_3$. The number of expected signal  events, $n_{\mathrm{sig}}$, is evaluated based on MC simulation assuming the cross section of the model under test multiplied by the signal strength $\mu$.

The significance of any deviation observed in the data with respect to the background-only expectation is quantified in terms of the local $p_0$ value. This is defined as the probability of fluctuations of the background-only expectation to produce an excess at least as large as the one observed. 
The largest deviation from the background model occurs in the $ZZ$ SR for a heavy scalar
with mass of 2.4~\TeV. The local significance of this deviation is $2.0~\sigma$ and
the corresponding global significance is less than $1~\sigma$.
No statistically significant excess is observed and upper exclusion limits
are placed on the cross section times branching ratio
for the production of heavy resonances decaying into diboson final states.
A correction to account for the branching ratio of $V$ decays into hadronic final states
is applied in the results below.
The limits are set with the CL$_{\text s}$ method \cite{Read:2002hq} using large 
sets of pseudo-experiments.

Limits on \sigbr\ are set in each combined diboson channel as a function of the resonance mass. The HVT models A and B with degenerate $W'$ and $Z'$ are used as benchmarks for the combined $WW+WZ$ signal region, and the bulk RS or heavy scalar models are used for the $WW+ZZ$ signal region. Figure~\ref{fig:WWWZ-limit} shows the observed limits on the production of a spin-1 vector triplet as a function of resonance mass in the $WW+WZ$ signal region. A spin-1 vector triplet with couplings predicted by the HVT model A (B) with $g_V = 1$ ($g_V = 3$) is excluded in the range $1.2<m(V')<3.1$ ($1.2<m(V')<3.5$)~\TeV, at the $95\%$ confidence level (CL). Figure~\ref{fig:WWZZG-limit} shows the observed limits on the production of a $\GKK$ as a function of $m(\GKK)$ in the $WW+ZZ$ signal region. Production of a ${\GKK}$ in the bulk RS model with $\kMPl=1$ is excluded in the range $1.3<m(\GKK)<1.6$~\TeV, at the $95\%$ CL.  Figure~\ref{fig:WWZZH-limit} shows the observed limits on the production of a new heavy scalar as a function of $m(\text{Scalar})$ in the $WW+ZZ$ signal region. 
Table~\ref{Res:tab:excl_mass} presents the resonance mass ranges excluded at the 95\% CL
in the various signal regions and signal models considered in the search.
In the search for heavy scalar particles, upper limits are set on \sigbr\ at the 95\% CL
with values of 9.7~fb at $m(\text{Scalar}) = 2$~\TeV{} and 3.5~fb
at $m(\text{Scalar}) = 3$~\TeV{}.

\begin{figure}[hpt!]
\begin{center}
  \subfloat[$WW+WZ$ signal region for HVT model]{
\includegraphics[scale=\scfigtwo]{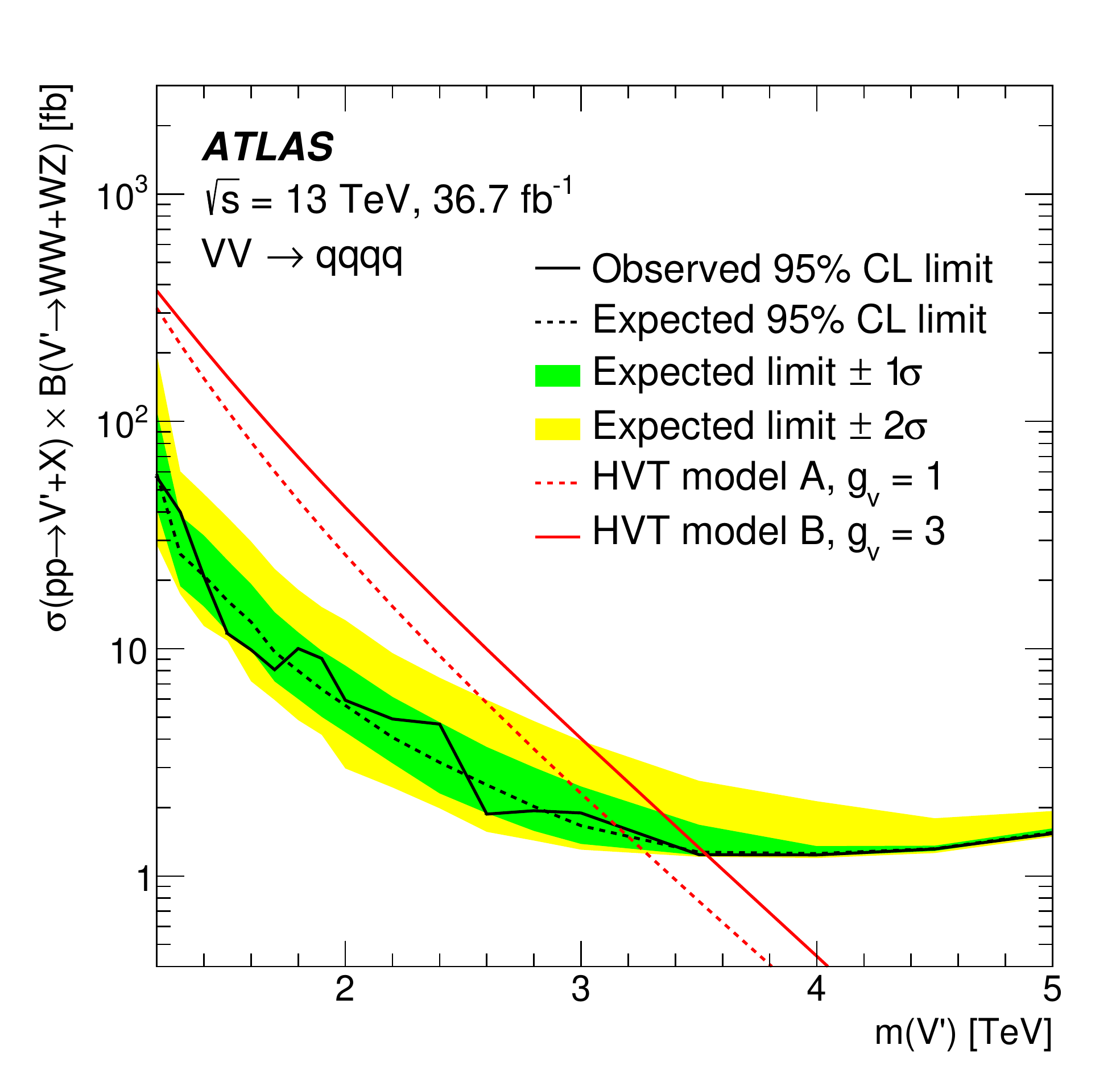}
\label{fig:WWWZ-limit}
} 
  \subfloat[$WW+ZZ$ signal region for bulk RS model]{
\includegraphics[scale=\scfigtwo]{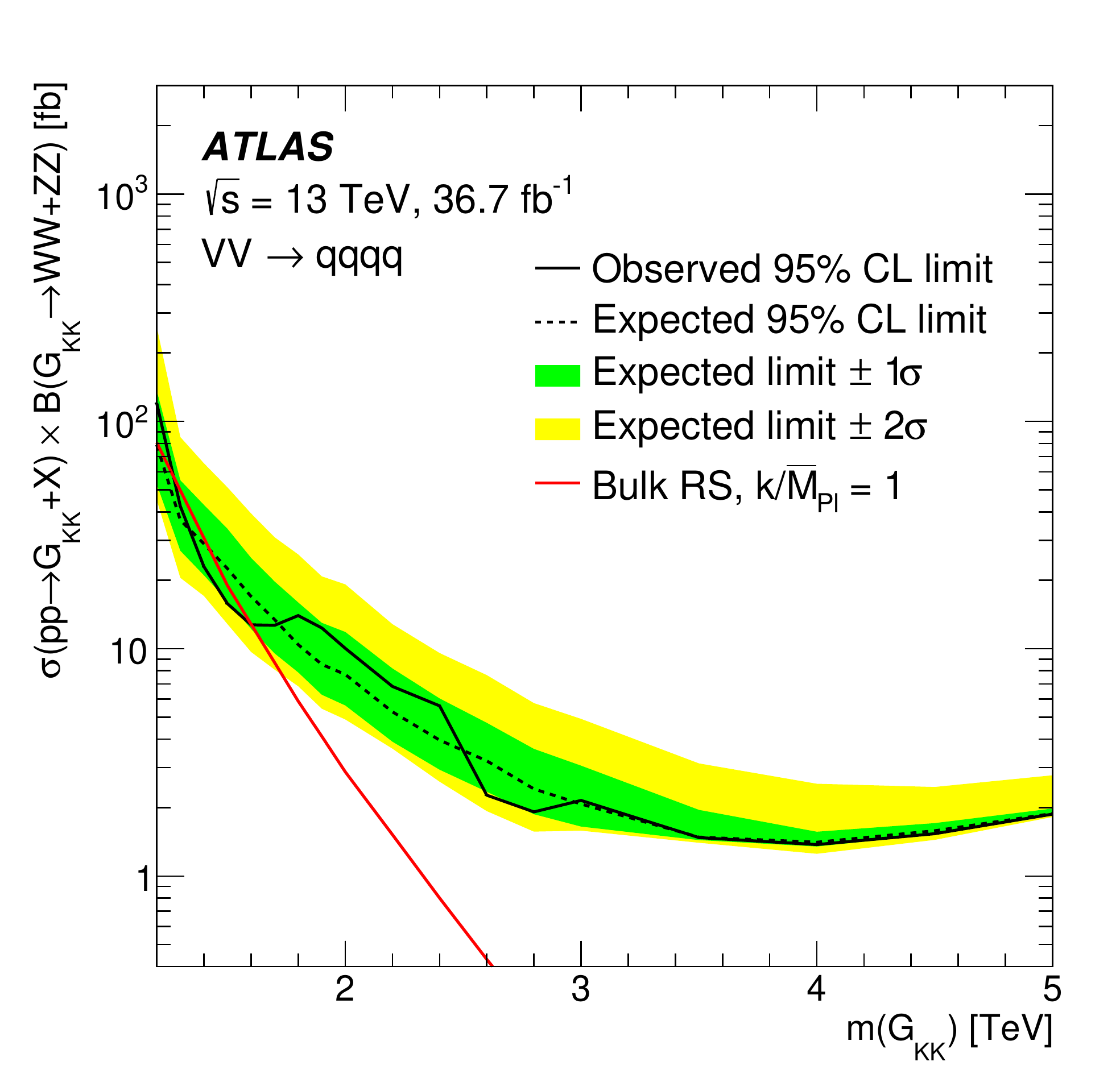}
\label{fig:WWZZG-limit}
} \\
  \subfloat[$WW+ZZ$ signal region for heavy scalar model]{
\includegraphics[scale=\scfigtwo]{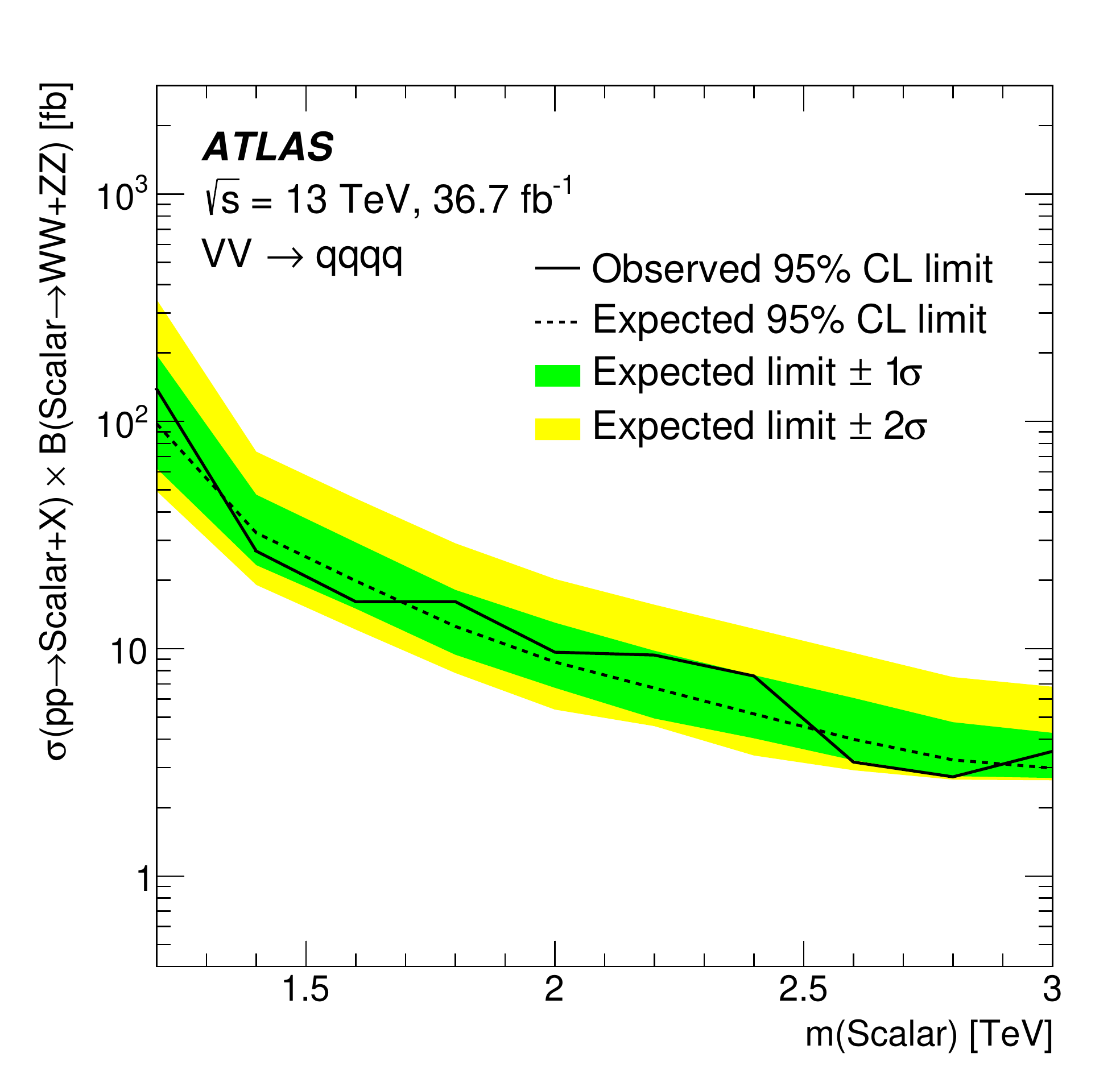}
\label{fig:WWZZH-limit}
}
\end{center}
\vspace{-20 pt}
\caption{Upper limits at the $95\%$ CL on the cross section times branching ratio for (a) $WW+WZ$ production as a function of $V'$ mass, (b) $WW+ZZ$ production as a function of $\GKK$ mass, and (c) $WW+ZZ$ production as a function of scalar mass. The predicted cross section times branching ratio is shown (a) as dashed and solid lines for the HVT models A with $g_V = 1$ and B with $g_V = 3$, respectively, and (b) as a solid line for the bulk RS model with $\kMPl=1$.
\label{fig:Data-limits}
}
\end{figure} 

\begin{table}[tpb!]
\caption{Observed excluded resonance masses (at 95\% CL) in the individual and combined signal regions for the HVT and bulk RS models.
 \label{Res:tab:excl_mass}
}
\begin{center}
\begin{tabular}{ccc}
\hline
\hline
Model   & Signal Region & Excluded mass range [$\TeV$]\\
\hline
                       & $WW$ & $1.20 - 2.20$ \\
HVT model A, $g_V = 1$ & $WZ$ & $1.20 - 3.00$ \\
                       & $WW+WZ$ & $1.20 - 3.10$ \\
\hline
                       & $WW$ & $1.20 - 2.80$ \\
HVT model B, $g_V = 3$ & $WZ$ & $1.20 - 3.30$ \\
                       & $WW+WZ$ & $1.20 - 3.50$ \\
\hline
                   & $WW$ & $1.30 - 1.45$ \\
Bulk RS, $\kMPl=1$ & $ZZ$ & none  \\
                   & $WW+ZZ$ & $1.30 - 1.60$ \\
\hline
\hline
\end{tabular}
\end{center}
\end{table}

\FloatBarrier

\section{Conclusions}
\label{sec:conclusions}

This Letter reports a search for massive resonances decaying via
$WW$, $WZ$ and $ZZ$ into hadrons
with \lumiAll~\ifb\ of $\sqrt{s}=13$~TeV $pp$ collisions collected at the LHC
with the ATLAS detector in 2015--2016.
The search takes advantage of the high branching ratio of hadronic decays
of the vector bosons and covers the resonance mass range between 1.2 and 5.0~\TeV.
In this kinematic range, the vector bosons are highly boosted and are 
reconstructed as single large-radius jets that are tagged by exploiting their two-body substructure. 
The invariant mass distribution of the two highest-\pt large-radius jets 
in each event is used to search for narrow resonance peaks over 
a smoothly falling background.
No significant excess of data is observed and limits are set on the cross section 
times branching ratio for diboson resonances at the 95\% confidence level.
In the case of the phenomenological HVT model A (model B) with $g_V = 1$ ($g_V = 3$), 
a spin-1 vector triplet is excluded for masses between 1.2 and 3.1 \TeV{}
(1.2 and 3.5 \TeV{}).
For the bulk RS model with $\kMPl = 1$, a spin-2 Kaluza--Klein graviton is excluded 
in the range between 1.3 and 1.6~\TeV. 
Upper limits on the production cross section times branching ratio for new heavy
scalar particles are set 
with values of 9.7 fb and 3.5 fb at scalar masses of 2~\TeV{} and 3~\TeV{}, respectively.


\section*{Acknowledgements}

We thank CERN for the very successful operation of the LHC, as well as the
support staff from our institutions without whom ATLAS could not be
operated efficiently.

We acknowledge the support of ANPCyT, Argentina; YerPhI, Armenia; ARC, Australia; BMWFW and FWF, Austria; ANAS, Azerbaijan; SSTC, Belarus; CNPq and FAPESP, Brazil; NSERC, NRC and CFI, Canada; CERN; CONICYT, Chile; CAS, MOST and NSFC, China; COLCIENCIAS, Colombia; MSMT CR, MPO CR and VSC CR, Czech Republic; DNRF and DNSRC, Denmark; IN2P3-CNRS, CEA-DRF/IRFU, France; SRNSF, Georgia; BMBF, HGF, and MPG, Germany; GSRT, Greece; RGC, Hong Kong SAR, China; ISF, I-CORE and Benoziyo Center, Israel; INFN, Italy; MEXT and JSPS, Japan; CNRST, Morocco; NWO, Netherlands; RCN, Norway; MNiSW and NCN, Poland; FCT, Portugal; MNE/IFA, Romania; MES of Russia and NRC KI, Russian Federation; JINR; MESTD, Serbia; MSSR, Slovakia; ARRS and MIZ\v{S}, Slovenia; DST/NRF, South Africa; MINECO, Spain; SRC and Wallenberg Foundation, Sweden; SERI, SNSF and Cantons of Bern and Geneva, Switzerland; MOST, Taiwan; TAEK, Turkey; STFC, United Kingdom; DOE and NSF, United States of America. In addition, individual groups and members have received support from BCKDF, the Canada Council, CANARIE, CRC, Compute Canada, FQRNT, and the Ontario Innovation Trust, Canada; EPLANET, ERC, ERDF, FP7, Horizon 2020 and Marie Sk{\l}odowska-Curie Actions, European Union; Investissements d'Avenir Labex and Idex, ANR, R{\'e}gion Auvergne and Fondation Partager le Savoir, France; DFG and AvH Foundation, Germany; Herakleitos, Thales and Aristeia programmes co-financed by EU-ESF and the Greek NSRF; BSF, GIF and Minerva, Israel; BRF, Norway; CERCA Programme Generalitat de Catalunya, Generalitat Valenciana, Spain; the Royal Society and Leverhulme Trust, United Kingdom.

The crucial computing support from all WLCG partners is acknowledged gratefully, in particular from CERN, the ATLAS Tier-1 facilities at TRIUMF (Canada), NDGF (Denmark, Norway, Sweden), CC-IN2P3 (France), KIT/GridKA (Germany), INFN-CNAF (Italy), NL-T1 (Netherlands), PIC (Spain), ASGC (Taiwan), RAL (UK) and BNL (USA), the Tier-2 facilities worldwide and large non-WLCG resource providers. Major contributors of computing resources are listed in Ref.~\cite{ATL-GEN-PUB-2016-002}.

\printbibliography


\newpage
\begin{flushleft}
{\Large The ATLAS Collaboration}

\bigskip

M.~Aaboud$^\textrm{\scriptsize 137d}$,
G.~Aad$^\textrm{\scriptsize 88}$,
B.~Abbott$^\textrm{\scriptsize 115}$,
O.~Abdinov$^\textrm{\scriptsize 12}$$^{,*}$,
B.~Abeloos$^\textrm{\scriptsize 119}$,
S.H.~Abidi$^\textrm{\scriptsize 161}$,
O.S.~AbouZeid$^\textrm{\scriptsize 139}$,
N.L.~Abraham$^\textrm{\scriptsize 151}$,
H.~Abramowicz$^\textrm{\scriptsize 155}$,
H.~Abreu$^\textrm{\scriptsize 154}$
% ...
\bigskip
\\
% ...
$^{12}$ Institute of Physics, Azerbaijan Academy of Sciences, Baku, Azerbaijan\\
% ...
$^{88}$ CPPM, Aix-Marseille Universit{\'e} and CNRS/IN2P3, Marseille, France\\
% ...
$^{115}$ Homer L. Dodge Department of Physics and Astronomy, University of Oklahoma, Norman OK, United States of America\\
% ...
$^{119}$ LAL, Univ. Paris-Sud, CNRS/IN2P3, Universit{\'e} Paris-Saclay, Orsay, France\\
% ...
$^{137}$ $^{(a)}$ Facult{\'e} des Sciences Ain Chock, R{\'e}seau Universitaire de Physique des Hautes Energies - Universit{\'e} Hassan II, Casablanca; $^{(b)}$ Centre National de l'Energie des Sciences Techniques Nucleaires, Rabat; $^{(c)}$ Facult{\'e} des Sciences Semlalia, Universit{\'e} Cadi Ayyad, LPHEA-Marrakech; $^{(d)}$ Facult{\'e} des Sciences, Universit{\'e} Mohamed Premier and LPTPM, Oujda; $^{(e)}$ Facult{\'e} des sciences, Universit{\'e} Mohammed V, Rabat, Morocco\\
% ...
$^{139}$ Santa Cruz Institute for Particle Physics, University of California Santa Cruz, Santa Cruz CA, United States of America\\
% ...
$^{151}$ Department of Physics and Astronomy, University of Sussex, Brighton, United Kingdom\\
% ...
$^{154}$ Department of Physics, Technion: Israel Institute of Technology, Haifa, Israel\\
$^{155}$ Raymond and Beverly Sackler School of Physics and Astronomy, Tel Aviv University, Tel Aviv, Israel\\
% ...
$^{161}$ Department of Physics, University of Toronto, Toronto ON, Canada\\
% ...
$^{a}$ Also at Department of Physics, King's College London, London, United Kingdom\\
$^{b}$ Also at Institute of Physics, Azerbaijan Academy of Sciences, Baku, Azerbaijan\\
$^{c}$ Also at Novosibirsk State University, Novosibirsk, Russia\\
$^{d}$ Also at TRIUMF, Vancouver BC, Canada\\
$^{e}$ Also at Department of Physics {\&} Astronomy, University of Louisville, Louisville, KY, United States of America\\
% ...
$^{*}$ Deceased
\end{flushleft}


\clearpage
\appendix

%

\end{document}